\documentclass{jaa}
\pdfoutput=1 
\usepackage{lastpage}
\usepackage{graphicx}	
\usepackage{amsmath}	
\usepackage{amssymb}	
\usepackage{adjustbox}
\usepackage{longtable}
\usepackage{multirow}
\usepackage{float}
\usepackage{natbib}
\usepackage{hyperref} 
\usepackage{subfigure}

\def\apjs{{\it Astrophys.~J.~Suppl.}}
\def\apj{{\it Astrophys.~J.}}
\def\aj{{\it Astronom.~J.}}
\def\apjl{{\it Astrophys.~J.~Lett.}}

\def\mnras{{\it Mon.~Not.~R.~Astron.~Soc.}}

\def\aap{{\it Astron.~Astrophys.}}

\begin{document}\sloppy

\title{
Study of the transient nature of classical Be stars using multi-epoch optical spectroscopy}


\author{Gourav Banerjee\textsuperscript{1,*}, Blesson Mathew\textsuperscript{1}, K. T. Paul\textsuperscript{1}, Annapurni Subramaniam\textsuperscript{2}, Anjusha Balan\textsuperscript{1}, Suman Bhattacharyya$^{1}$, R. Anusha$^{3}$, Deeja Moosa$^{1}$, C S Dheeraj$^{1}$, Aleeda Charly$^{1}$ and Megha Raghu\textsuperscript{1}}

\affilOne{\textsuperscript{1}Department of Physics and Electronics, CHRIST (Deemed to be University), Hosur Main Road, Bengaluru, India\\}
\affilTwo{\textsuperscript{2}Indian Institute of Astrophysics, Koramangala, Bengaluru, India\\}
\affilThree{\textsuperscript{3}Department of Physics and Astronomy, University of Western Ontario, London, ON N6A 3K7, Canada}


\twocolumn[{

\maketitle

\corres{gourav.banerjee@res.christuniversity.in}


\begin{abstract}
Variability is a commonly observed property of classical Be stars (CBe) stars. In extreme cases, complete disappearance of the H$\alpha$ emission line occurs, indicating a disc-less state in CBe stars. The disc-loss and reappearing phases can be identified by studying the H$\alpha$ line profiles of CBe stars on a regular basis. In this paper, we present the study of a set of selected 9 bright CBe stars, in the wavelength range of 6200 - 6700 \AA, to better understand their disc transient nature through continuous monitoring of their H$\alpha$ line profile variations for 5 consecutive years (2015 -- 2019). Based on our observations, we suggest that 4 of the program stars (HD 4180, HD 142926, HD 164447 and HD 171780) are possibly undergoing disc-loss episodes, whereas one other star (HD 23302) might be passing through disc formation phase. The remaining 4 stars (HD 237056, HD 33357, HD 38708 and HD 60855) have shown signs of hosting a stable disc in recent epochs. Through visual inspection of the overall variation observed in the H$\alpha$ EW for these stars, we classified them into groups of growing, stable and dissipating discs, respectively. Moreover, our comparative analysis using the BeSS database points out that the star HD 60855 has passed through a disc-less episode in 2008, with its disc formation happening probably over a timescale of only 2 months, between January and March 2008.
\end{abstract}


}]


\doinum{12.3456/s78910-011-012-3}
\artcitid{\#\#\#\#}
\volnum{000}
\year{0000}
\pgrange{1--}
\setcounter{page}{1}
\lp{1}

\section{Introduction}
Classical Be (CBe) stars are main sequence or evolved stars, belonging to luminosity classes III-V and having masses and radii ranging between $M_\star$ $\sim$ $3.6-20$ $M_\odot$, and $R_\star$ $\sim$ $2.7-15$ $R_\odot$ (\cite{2000Cox}). As defined by \cite{1987Collins}, a CBe star is “a non-supergiant B star whose spectrum has, or had at some time, one or more Balmer lines in emission”. They can be easily distinguished from normal B-type stars by the presence of recombination emission lines of different elements such as hydrogen, iron, oxygen, helium, calcium, silicon, etc \cite[e.g.][]{2021Banerjee, 2018Shokry, 2017Aguayo, 2012Paul, 2011Mathew, 1996Hanuschik, 1988Andrillat, 1982AndrillatF} in their spectra, in addition to an infrared excess noticed in their continuum \citep{1974Gehrz,1977Hartmann}. These characteristics are indicative of the presence of a circumstellar disc \citep{2013Rivinius}. Being an equatorial, geometrically thin, gaseous, decretion disc in nature, it orbits the central star in Keplerian rotation \citep{2006Carciofi, 2007Meilland}.

\cite{1931Struve} first proposed that the ejection of stellar material is responsible for the formation of a circumstellar disc in a CBe star. Presently, the physical model which best describes these discs is the viscous decretion disc (VDD) model \citep{2012Carciofi,1991Lee}. However, the disc formation mechanism in CBe stars - the ‘Be phenomenon’ - is still poorly understood. It is well understood that spectral analysis of different emission lines seen in CBe star spectra provides a wealth of information about the geometry and kinematics of the gaseous disc and several properties of the central star itself (e.g. \citealt {2021Banerjee, 2019Klement, 2018Mennickent, 2013Barnsley, 2012bMathew, 2011Mathew, 2006Granada, 1992Dachs, 1990Dachs, 1984Dachs, 1987Hanuschik, 1985Chalabaev, 1976Polidan}). Hence, a considerable amount of studies have been carried out focusing on this aspect to better understand the ‘Be phenomenon’.

CBe stars usually exhibit variability in spectral line profiles. They show in their spectra either short-term variations occurring on timescales of hours to months  \cite[e.g.][]{2017Paul, 2003Porter, 1996Sterken, 1986Penrod, 1982Baade} or long-term variability which occurs on timescales of years to decades \citep{1994Mennickent, 1991Mennickentb}. Non-radial pulsation (NRP) explains significant line profile variability (LPV) on timescales ranging from hours to days \citep{1982Baade}. \cite{2003Porter} reported that NRP can explain the observed LPV for early-type CBe stars in around 80\% of the cases. Moreover, binarity effects have also been found to be an important source of intermediate-period disc variability (\cite{2018Panoglou}, and references therein). The extreme case of such variability is the disappearance of the H$\alpha$ emission line, indicative of a disc-less state in CBe stars. The spectrum then looks like that of a normal B-type star with photospheric absorption lines. A well-studied example is the disc-less episode of X Persei \citep{1992Fabregat, 1991Norton}. Infrared studies of this star were also performed by \cite{2012aMathew, 2013Mathew}.

It is possible to have a change in line profile shape and intensity by altering the disc density values, as demonstrated in the study of \cite{2010Silaj}. Interestingly, observations during such a disc-less state can be used to estimate the stellar parameters such as spectral type and luminosity \citep{1992Fabregat}. The disc-loss and reappearing phases of CBe stars can be identified by studying their H$\alpha$ line profiles on a regular basis.

During our previous study of field CBe stars using the Himalayan Chandra Telescope (HCT), we observed that 22 among our 118 program stars showed H$\alpha$ in absorption \citep{2021Banerjee}. However, after correcting for the underlying stellar photosphere, it was detected that H$\alpha$ is present indeed in absorption in case of only one among these 22 stars, i.e., in the case of HD 60855. We then looked into the literature to check previous studies devoted to understand the disc-loss and reappearing phases of CBe stars from the analysis of the H$\alpha$ emission line profiles on a regular basis. It was found that several studies exist which reported the disc-less state of a few CBe stars \cite[few recent works are][]{2021Richardson, 2021Marr, 2021Cochetti, 2021Ghoreyshi, 2019Klement, 2011Mathew}. However, we noticed a scarcity in literature about any work which discusses the disc formation and dissipation in CBe stars in terms of continuous changes in H$\alpha$ line profile. Motivated by these facts we decided to perform a dedicated study to understand the disc transient nature of CBe stars through continuous monitoring of their H$\alpha$ line profile variations. Moreover, \cite{2021Ghoreyshi} found that early type CBe stars such as $\omega$ CMa  possess more unstable disc than late types, in agreement with the results of \cite{2018Labadie-Bartz}. These results inspired us further to perform our present work.

Furthermore, some recent studies have also identified several new emission-line stars of different types using photometry and spectroscopy. While \cite{2021Anusha} identified 159 Classical Ae (CAe) stars, which are regarded as the lower mass analogs of CBe stars, \cite{2021Li} detected 12 (6 of which are new detections) Oe stars, thought to be the hotter counterpart of CBe stars \citep{2013Rivinius}. In a separate study, \cite{2021Bhattacharyya} identified 98 ‘Transition Phase’ (TP) candidates: rare stars evolving from pre-main sequence to main sequence phase. Another interesting study by \cite{2021Madhu} have even identified 13 new CBe stars in open clusters older than 100 Myr. Moreover, effects of metallicity in CBe stars has already been noticed in several previous studies \cite[e.g.][]{1999Maeder, 2004Keller, 2006Martayan, 2007Martayanb, 2007Martayan}. Hence, further studies of CBe stars in different metallicity environments (such as clusters and fields) might even provide clues about the disc nature of these different types of emission-line stars.

In the present paper, we present the results of our study of a set of carefully selected 9 bright Galactic CBe stars which have shown H$\alpha$ line in complete absorption at least once in literature, indicating that these stars have passed through a disc-less phase at least once in their lifetime. We considered the star HD 60855 which exhibited H$\alpha$ line in complete absorption during our previous observation using the HCT \citep{2021Banerjee} for this study since it is a bright source. The remaining 8 stars are selected through literature survey and considering the fact that they are observable with the facility we used for our observations (discussed in Sect. \ref{Section2:Obs}). We studied the transient nature of these nine CBe stars over a period of 5 years (2015 - 2019) by analyzing the continuous changes in H$\alpha$ line profile shown by them. Our study will help in the modelling of circumstellar discs of CBe stars and thus provide a better understanding about the ‘Be phenomenon' in CBe stars. Sect. \ref{Section2:Obs} discusses the observations performed and about the datasets used for this study. The results obtained from the analysis of the data are discussed in Sect. \ref{Section3}. The prominent results from our study are summarised in Sect. \ref{Section4}.

\section{Observations}
\label{Section2:Obs}
We obtained 140 medium resolution spectra of selected 9 bright CBe stars (V from 3.7 to 8.9) using the Universal Astronomical Grating Spectrograph (UAGS) instrument fitted with the 1.0--m reflecting telescope situated at the Vainu Bappu Observatory (VBO), Kavalur, Tamil Nadu, India. VBO is operated by the Indian Institute of Astrophysics, Bengaluru\footnote{$(http://www.iiap.res.in/)$}. These stars were selected since they are bright enough for observing with the 1.0--m telescope. The CCD used for imaging consists of 1024 $\times$ 1024 pixels with a size of 24 $\mu$m, where the central 1024 $\times$ 300 pixels were used for spectroscopy. Continuous monitoring of these stars was carried out since June 2017. We used the Bausch and Lomb 1800 lines/mm grating, which in combination with the slit gives 1 \AA~resolution at H$\alpha$. The spectra were obtained in the range of 6200 - 6700 \AA, at settings particularly centered at H$\alpha$ line. This is done for studying the transient nature of CBe stars by investigating the continuous changes observed in their H$\alpha$ line profile. Dome flats taken with halogen lamps were used for flat-fielding the images. Bias subtraction, flat-field correction and spectral extraction were performed with standard IRAF tasks. FeNe lamp spectra were taken along with the object spectrum for wavelength calibration. All the extracted raw spectra were wavelength calibrated and continuum normalized with IRAF tasks. Multiple exposures of 10 - 40 minutes were required to receive a good signal. The log of our observations is presented in Table \ref{table1}.

Moreover, we found that 3 among these 9 stars were observed by our research team using the same setup since 2015. So we retrieved the data for those stars which were previously observed during 2015 -- 2016. This increased the data volume for our present study.

\section{Results}
\label{Section3}
In this section we present the analysis of 9 CBe stars. They belong to spectral types B0 -- B8. Out of 9 stars, 4 are earlier than the B5 type, whereas others are later than B5. It is also noted from the literature that one among these 9 stars, HD 142926 (4 Her), is a shell star.

Prominent spectral lines identified in the spectra of our sample of CBe stars are highlighted in Sect. 3.1. Sect. 3.2 describes the major spectral features observed in each star along with our interpretations. A comparative study with the data available from the BeSS database is presented in Sect. 3.3. Sect. 3.4 discusses the overall variation of the H$\alpha$ equivalent width (EW) for the program stars as observed based on our data. In the last section, a focused discussion is done about those particular stars which have either undergone or undergoing disc-loss episodes.

\subsection{Prominent spectral lines identified in our CBe sample}
\begin{figure*}
\centering
   \includegraphics[height=170mm, width=160mm]{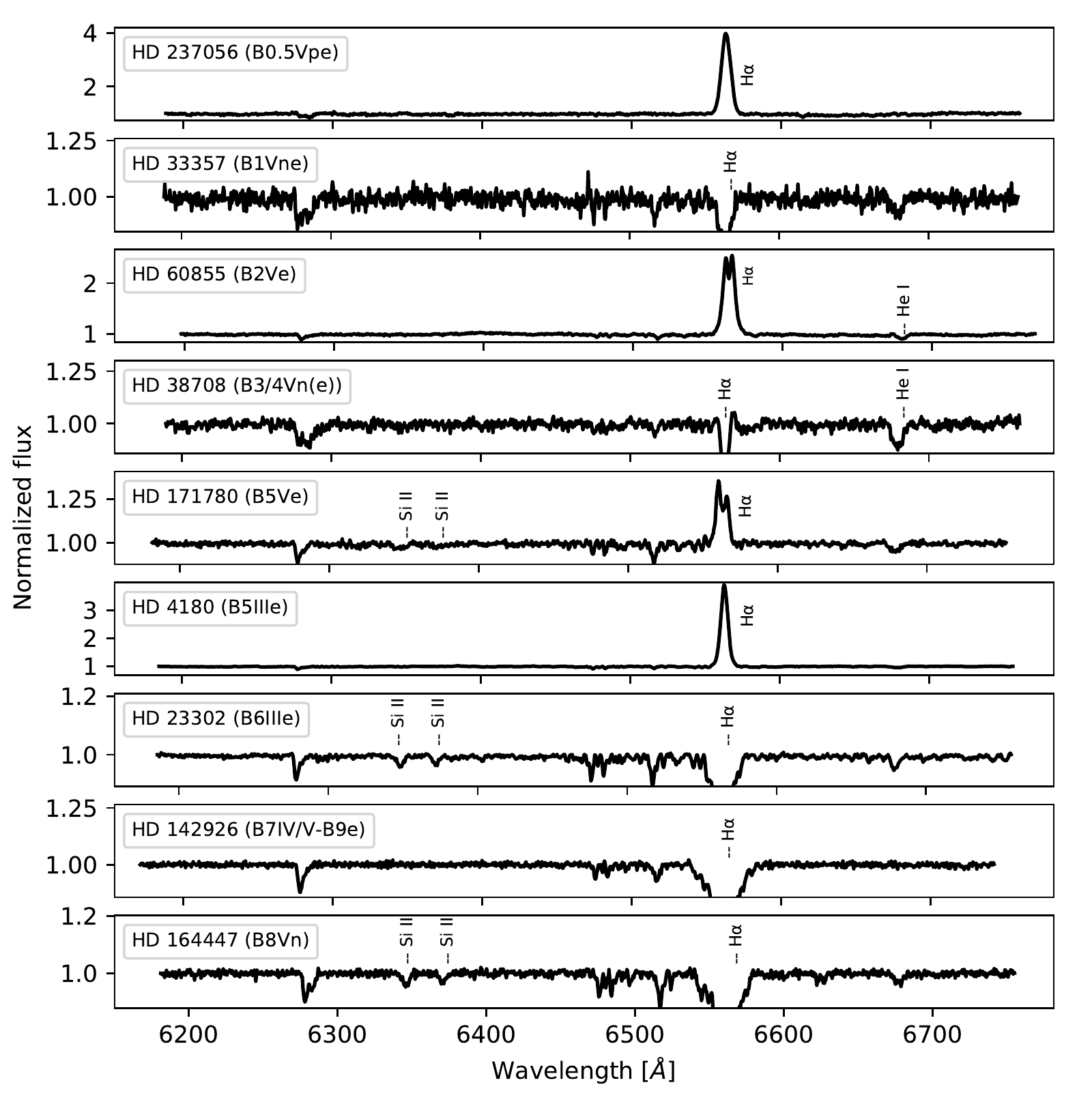}
\caption{Representative spectra of all 9 of our sample stars showing different spectral features in the wavelength range of 6200 - 6700 {\AA}. It is found that while HD 237056 (B0) and HD 4180 (B5) show H$\alpha$ in single peak emission, HD 60855 (B2) and HD 171780 (B5) exhibit H$\alpha$ in double peak emission. While V $<$ R in case of HD 60855, HD 171780 shows V $>$ R. H$\alpha$ is present in absorption for the rest 5 stars. It  that the He{\sc i} 6678 \AA~absorption line is visible in case of the early type stars HD 60855 and HD 38708, whereas . Si{\sc ii} 6347, 6371 \AA~lines appear in absorption in case of HD 171780, HD 23302 and HD 164447, respectively.}
\label{fig1Rep}
\end{figure*}

At first, we identified major spectral features observed in each spectrum for all 9 of our program stars. We found that the H$\alpha$ line is visible in either single or double-peak emission in all spectra for 4 stars, namely HD 4180, HD 237056, HD 60855 and HD 171780. While HD 4180 and HD 237056 show H$\alpha$ in single peak emission, double-peak emission of H$\alpha$ is noticed in HD 60855 and HD 171780, respectively. H$\alpha$ appears in absorption for the remaining 5 stars. Emission in absorption ({\it eia}) profile for H$\alpha$ is also observed for 3 stars: HD 164447 in 8 cases, HD 23302 in 5 occasions and HD 171780 only once (November 4, 2019). Apart from H$\alpha$, we also detected absorption lines of He{\sc i} 6678 \AA, Si{\sc ii} 6347, 6371 \AA~occasionally for our sample stars. A set of representative spectra for all 9 stars showing different spectral features in the wavelength range of 6200 - 6700 {\AA} is presented in Fig. \ref{fig1Rep}. The average signal-to-noise ratio (SNR) of the spectra of every CBe star used for this study is usually greater than 90.

We observed a little to moderate variations of H$\alpha$ emission EW for all of our stars. The H$\alpha$ EW for our sample ranges from -0.6 (for HD 33357 on February 24, 2017) to -33.6 Å (for HD 4180 on September 17, 2015). In general, the variation is within 20\% for all the stars, while comparing the multi-epoch spectra of one star. A gradual increase of H$\alpha$ EW from -8 to -9.4 \AA~was exhibited by the star HD 60855 with minor fluctuations every time. On the contrary, a gradual decrease of H$\alpha$ EW was found for 3 stars, namely HD 4180 (from -33.6 to -24.5 \AA), HD 164447 (from -8.5 to -5.7 \AA) and HD 171780 (from -15.7 to -8.4 \AA), respectively. The H$\alpha$ profile observed for each star is further described in Sect. 3.2.

Since in-filling of photospheric absorption lines occur, visual inspection is not reliable in discerning between a star showing weak emission and another one exhibiting no emission at all. Hence, we calculated the H$\alpha$ EW for all our stars first and then measured the absorption component at the H$\alpha$ line from the synthetic spectra using models of stellar atmospheres \citep{1993Kurucz} for each spectral type. We estimated the corrected H$\alpha$ EW for our program stars following the method described in Sect. 3.4.1 of \cite{2021Banerjee}. Columns 9 and 10 of Table \ref{table1} presents the measured and corrected H$\alpha$ EW for all the 9 stars on every occasion.

We performed the error measurements of H$\alpha$ EWs using two different methods. At first, we measured the area under the curve for every case for individual stars using IRAF tasks. Then, we used a Voigt profile for profile fitting and measured the EW values similarly for every case. Next, we calculated the mean values for both these methods in case of individual stars. The standard deviation found from these measurements is considered as the error (shown in Table \ref{table2}).

Apart from H$\alpha$, He{\sc i} 6678 \AA~absorption line is observed for 4 stars, namely HD 237056, HD 33357, HD 38708 and HD 60855. All of them belong within B3 spectral type, which is in agreement with \cite{2021Banerjee}. While HD 38708 shows He{\sc i} 6678 \AA~ line in absorption each time, it is found occasionally in case of the remaining 3 stars. Furthermore, Si{\sc ii} 6347 and 6371 \AA~lines are noticed in absorption in 5 out of 9 stars, except for HD 4180, HD 237056, HD 33357 and HD 38708. Among them, 4 are late-type CBe stars belonging to spectral types B5 and beyond, whereas only HD 60885 is of B2 type. Hence, our result suggests that Si{\sc ii} 6347, 6371 \AA~absorption lines might be common in late type CBe stars with spectral types B5 and beyond.

\subsection{Spectral features observed in each of our program stars}
This subsection describes the variation of H$\alpha$ line profile as observed by us in each of the program stars. The measured and corrected H$\alpha$ EW for all 9 stars found on every date of our observations are shown in columns 9 and 10 of Table \ref{table1}. Likewise, the details of other spectral features observed on all dates of observation for individual stars is presented in columns 5, 6 and 7 of Table \ref{table2}. Among the 9 stars studied in this work, 6 are reported as binaries in the literature.

\subsubsection{\textbf{HD 4180 (Omi Cas)}}
HD 4180 is a well-studied, bright CBe star of spectral type B5IIIe. It was classified and cataloged as a CBe star by \cite{1982Jaschek}. It has shown remarkable H$\alpha$ line profile changes which is summarized by \cite{1972Peton}. \cite{1979Hubert-Delplace} reported that this star did not show any H$\alpha$ emission from 1953 to 1959. Later, \cite{1978Slettebak} and \cite{1982AndrillatF} observed it to show H$\alpha$ in emission.

When observed using the Himalayan Chandra Telescope (HCT) on December 03, 2007, we found its corrected H$\alpha$ EW to be -39.6 \AA~ \citep{2021Banerjee}. Then until 2018, we could obtain 5 spectra for this star centering at H$\alpha$ using the UAGS facility at Kavalur. Among these 5, one spectrum is retrieved from the observations done in 2015 using the same setup. Fig. \ref{fig2OmiCas} shows the H$\alpha$ line profile variation observed for HD 4180 over a timescale of 28 months. Following Table \ref{table1} it is seen that the H$\alpha$ line is visible in emission on every occasion of our observation. Overall, we found that the corrected H$\alpha$ EW gradually decreased by over 8 \AA~ for this star. This suggests that HD 4180 might be undergoing disc-loss since December 2007, accompanied by moderate variations in H$\alpha$ emission strength.

\subsubsection{\textbf{HD 237056 (BD+57 681)}} 
Classified as a CBe star by \cite{1982Jaschek}, HD 237056 belongs to spectral type B0.5V \citep{1999Steele}. H$\alpha$ was found to be in absorption for this star by \cite{1999Steele}. 

\begin{figure*}
  \centering
  \begin{minipage}[b]{0.4\textwidth}
    \includegraphics[height=90mm, width=60mm]{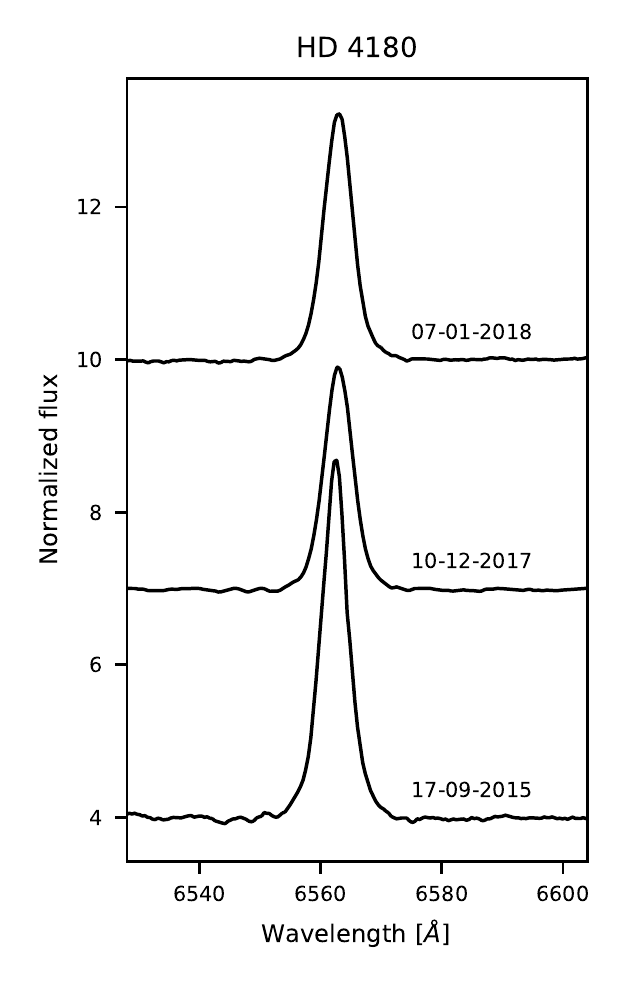}
  \end{minipage}
  \hspace{1em}
  \begin{minipage}[b]{0.4\textwidth}
    \includegraphics[height=90mm, width=60mm]{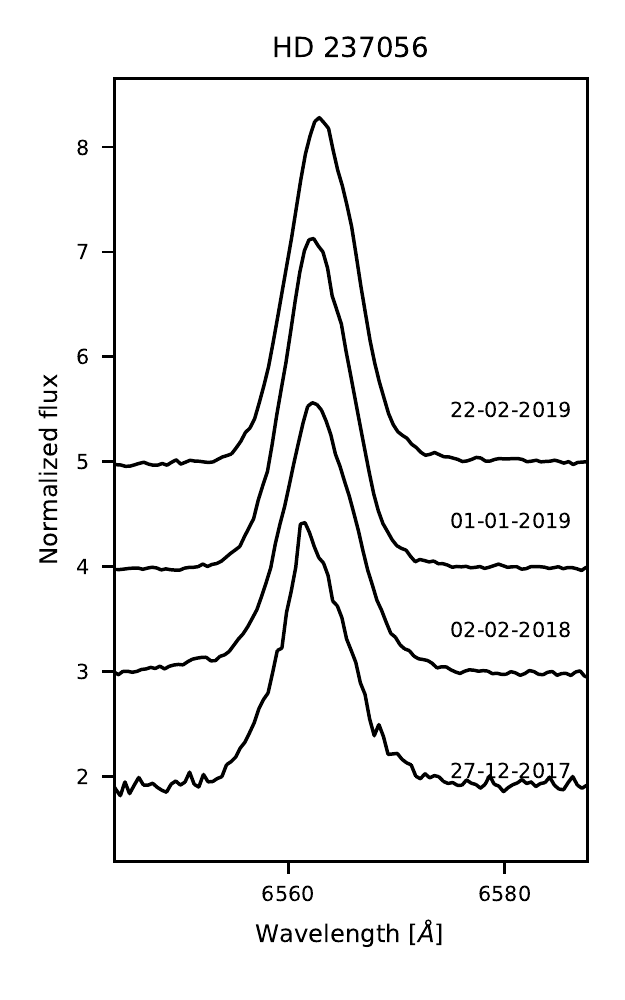}
  \end{minipage}
  \begin{minipage}[b]{0.4\textwidth}
    \includegraphics[height=90mm, width=60mm]{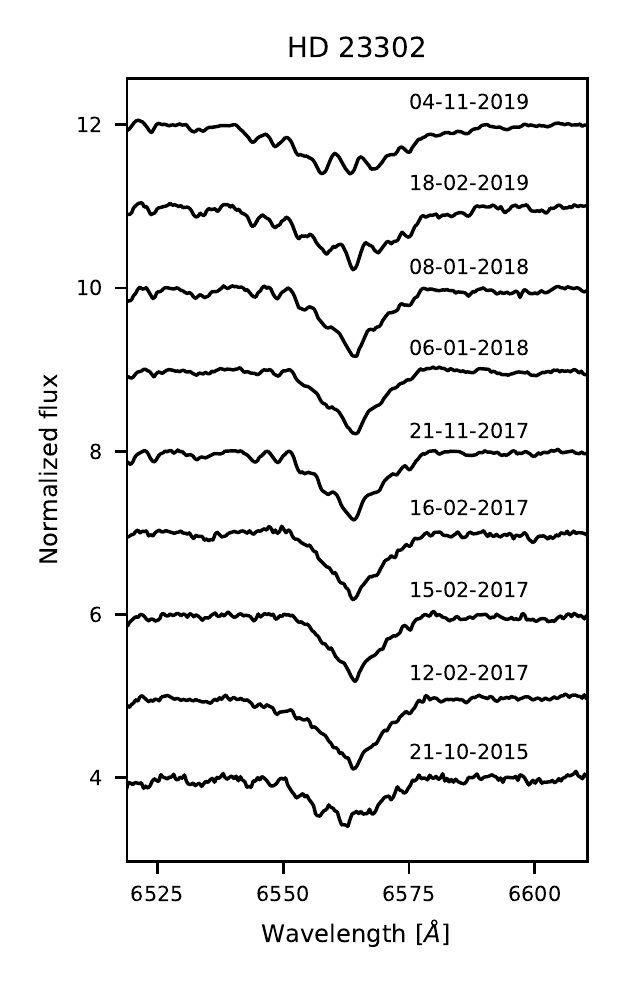}
  \end{minipage}
  \hspace{1em}
  \begin{minipage}[b]{0.4\textwidth}
    \includegraphics[height=90mm, width=60mm]{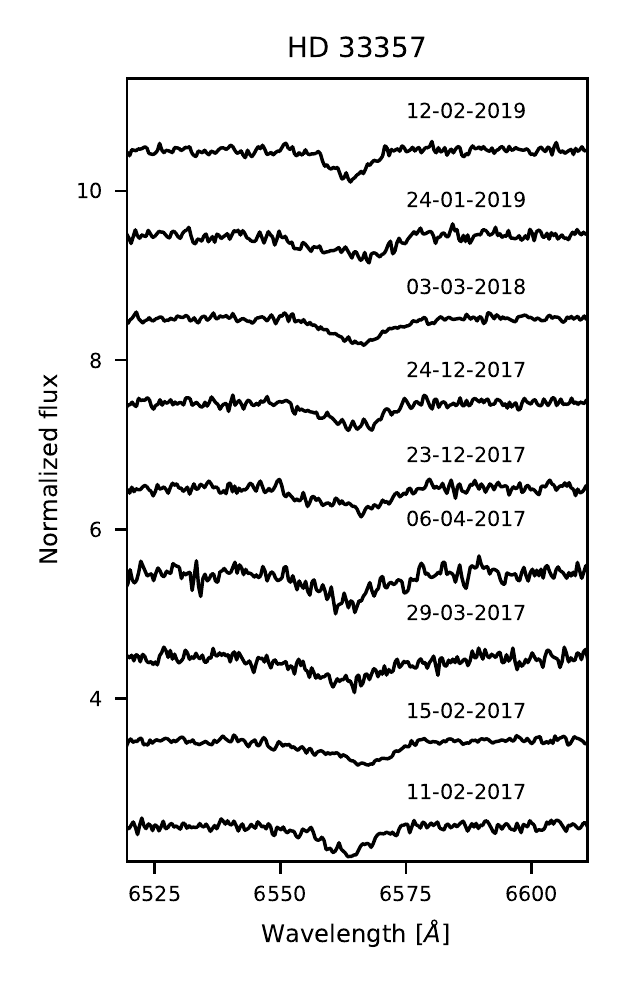}
  \end{minipage}
 \caption{H$\alpha$ line profile variation observed for HD 4180, HD 237056, HD 23302 and HD 33357 over a timescale of 28, 14, 49 and 24 months, respectively.}
\label{fig2OmiCas}
\end{figure*}

Using HCT, we observed HD 237056 on December 27, 2007 and its corrected H$\alpha$ EW was found to be -2.8 \AA~\citep{2021Banerjee}, which is weak. We then obtained 7 spectra for this star from Kavalur. Not much variation is noticed in the H$\alpha$ emission EW during our observations. From Table \ref{table1}, it is visible that the H$\alpha$ EW has changed by an amount of around 6 \AA, decreasing within two months and then increasing.  Fig. \ref{fig2OmiCas} presents the H$\alpha$ line profile variation observed for HD 237056 over a timescale of 14 months. Our results, thus indicate that this star's disc has formed considerably since 2007, whereas presently it is possibly hosting a stable disc. Moreover, this might be one of the state with large H$\alpha$ EW for this star.

\subsubsection{\textbf{HD 23302 (17 Tau)}}
Identified as a CBe star in the survey of \cite{1925Merrill} and later cataloged by \cite{1982Jaschek}, HD 23302 is a member of the Pleiades (M 45) cluster. \cite{1979Hubert-Delplace} observed it to exhibit H$\alpha$ always in absorption from 1955 to 1976.

We observed HD 23302 on December 20, 2008, using HCT and H$\alpha$ exhibited emission with EW of -7.5 \AA~\citep{2021Banerjee}. Through the Kavalur facility, we took 22 spectra for this star. Fig. \ref{fig2OmiCas} shows the H$\alpha$ line profile variation observed for HD 23302 over a timescale of 49 months. H$\alpha$ was apparently present in absorption most of the time. However, it is visible from Table \ref{table1} that HD 23302 is showing H$\alpha$ emission completely below the continuum until February 18, 2019. Then onward, H$\alpha$ exhibited emission in absorption profile having double-peak, with minor variations in EW. An overall H$\alpha$ EW variation of 4.2 \AA~is detected for this star during our observation period, while the EW remained above -7 \AA~since February 18, 2019. Such observations indicate that this star might be going through an episode of disc formation in the current epochs.

\subsubsection{\textbf{HD 33357 (SX Aur)}}
HD 33357 is an eclipsing binary with components of spectral types B3V and B5V \citep{2013Avvakumova}. It was also listed in the CBe star catalog of \cite{1982Jaschek}. However, from the analysis of IUE data, \citet{1985Peters} found no evidence for a circumstellar envelope.

Observation by \cite{2021Banerjee} using HCT on January 06, 2009 showed H$\alpha$ to be present in weak emission (corrected EW was -0.6 \AA). Using the Kavalur facility, we could obtain 15 spectra for this star. Fig. \ref{fig2OmiCas} presents the H$\alpha$ line profile (apparently present in absorption) variation observed for HD 33357 over a timescale of 24 months. Interesting variability of H$\alpha$ EW is detected for this star, as evident from Table \ref{table1}. After correcting for the underlying stellar photosphere, it was found that H$\alpha$ emission exists below the continuum having EW always $\leq$ -3.1 \AA. This indicates that HD 33357 might be a weak emitter by nature, possessing a stable disc in current epochs.

\subsubsection{\textbf{HD 38708 (V438 Aur)}} HD 38708 was cataloged as a B3pe shell star by \cite{1982Jaschek} and \cite{1997Kohoutek}.

Through HCT, we observed this star on December 02, 2008 and H$\alpha$ was present in weak emission (having corrected EW of -2.3 \AA) \citep{2021Banerjee}. Using the Kavalur facility, we took a total of 31 spectra of this star. Fig. \ref{fig3HD38708} presents the H$\alpha$ line profile variation observed for HD 38708 over a timescale of 36 months. H$\alpha$ was noticed in absorption which gradually became more intense since January 15, 2018. This may point out that this star might be moving towards a disc-loss phase. However, it was detected that weak H$\alpha$ emission exists below the continuum in every instance ($\leq$ -4.6 \AA) as shown in Table \ref{table1}. Our study thus suggests that HD 38708 can be a weak emitter by nature.

\subsubsection{\textbf{HD 60855 (V378 Pup)}}
listed in the H$\alpha$ emission-line star catalog of \cite{1943Merrill}, HD 60855 was cataloged as a B2Ve star by \cite{1982Jaschek}.

Interestingly, our observation of this star on January 11, 2008 with HCT showed H$\alpha$ in absorption having an EW of 4.9 \AA~\citep{2021Banerjee}. This was the first detection of a disc-less state exhibited by HD 60855. Then we took 22 spectra of it from Kavalur. H$\alpha$ is found to exhibit double-peak emission profile in every case with corrected EW ranging between -8 (March 27, 2016) to -9.4 \AA~(May 08, 2017). Although minor variations of H$\alpha$ EW is detected each time, the presence of a stable disc orbiting this star during the present decade is confirmed by our observations. However, our study using multi-epoch data also reveals that HD 60855 has passed through a disc-less episode during the recent past. The disc was not present when we observed it in 2008. Fig. \ref{fig3HD38708} presents the H$\alpha$ line profile variation was observed for HD 60855 over a timescale of 15 months.

\subsubsection{\textbf{HD 142926 (4 Her)}}
HD 142926 was recognized as a CBe star by \cite{1939Heard} and \cite{1940Mohler} and later cataloged by \cite{1982Jaschek}. \cite{1971Hubert} found that this star resembled the spectrum of a normal B type star from 1953 to 1963, with no visible H$\alpha$ emission. Then \citep{1994Koubsky} identified two periods of different duration when the H$\alpha$ emission was absent (1948-1963, 1987-1991) for this star.

A total of 16 spectra of HD 142926 were obtained using the Kavalur facility by us. Fig. \ref{fig3HD38708} presents the H$\alpha$ line profile variation observed for this star over a timescale of 24 months. Table \ref{table1} reveals that H$\alpha$ emission exists below the continuum with an overall decrease of EW of around 3.6 \AA. This is similar to the decrease found between 1997-1999 by \cite{2006Rivinius}, after a gap of about 20 years. Hence, our results can be suggestive of the fact that HD 142926 might be losing its disc gradually.

\subsubsection{\textbf{HD 164447 (V974 Her)}}
HD 164447 was listed by \cite{1982Jaschek} as a CBe star of spectral type B8Vne. However, \cite{1969Jaschek} did not detect any emission for this star in April 1967. Moreover, \cite{1976Svolopoulos} also did not notice any emission in the Balmer region in 1971. 

We obtained 14 spectra of this star from Kavalur and H$\alpha$ appeared to be in absorption in most cases. Fig. \ref{fig3HD38708} which presents the H$\alpha$ line profile variation observed for HD 164447 over a timescale of 43 months. However, looking into Table \ref{table1} it is found that H$\alpha$ emission exists below the continuum and shows an overall decrease of EW of over 3 \AA. Moreover, double-peak H$\alpha$ is visible in emission in absorption profile on all 8 occasions in 2016. Hence, our study suggests that the disc of HD 164447 might be dissipating away in recent epochs.

\subsubsection{\textbf{HD 171780 (HR 6984)}}
HD 171780 was listed as a CBe star in the catalog of \cite{1982Jaschek}. \cite{1993Jaschek} observed this star when it was in its B-phase. Subsequently, \cite{1998Moujtahid} mentioned it to be a B - Be variable star of spectral type B5Vne with moderate emission.

\begin{figure*}
  \centering
  \begin{minipage}[b]{0.4\textwidth}
    \includegraphics[height=78mm, width=50mm]{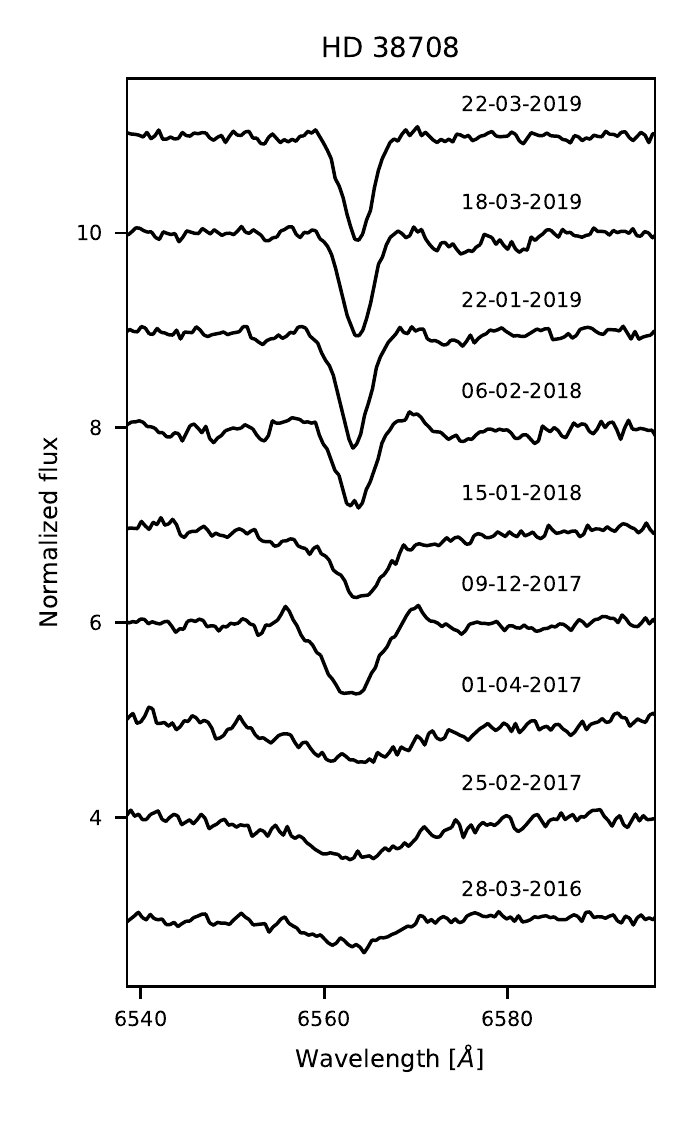}
  \end{minipage}
  \hspace{1em}
  \begin{minipage}[b]{0.4\textwidth}
    \includegraphics[height=78mm, width=50mm]{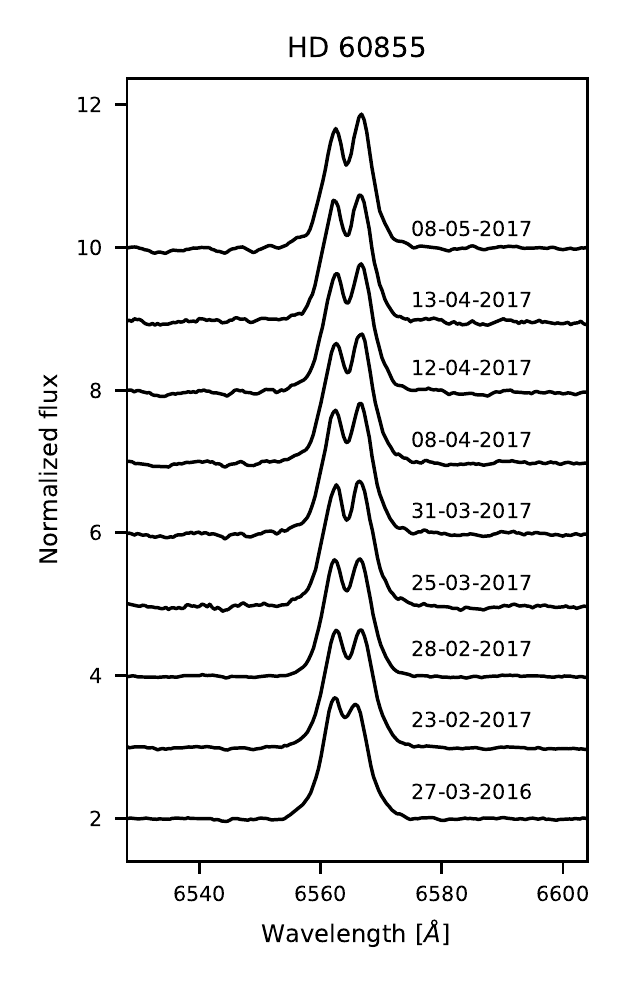}
  \end{minipage}
  \begin{minipage}[b]{0.4\textwidth}
    \includegraphics[height=78mm, width=50mm]{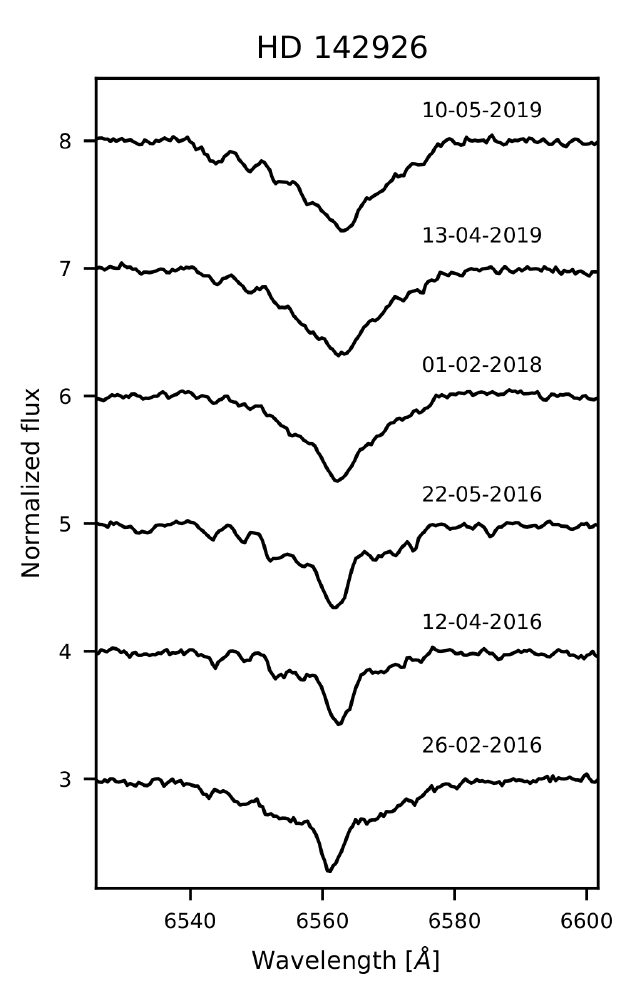}
  \end{minipage}
  \hspace{1em}
  \begin{minipage}[b]{0.4\textwidth}
    \includegraphics[height=78mm, width=50mm]{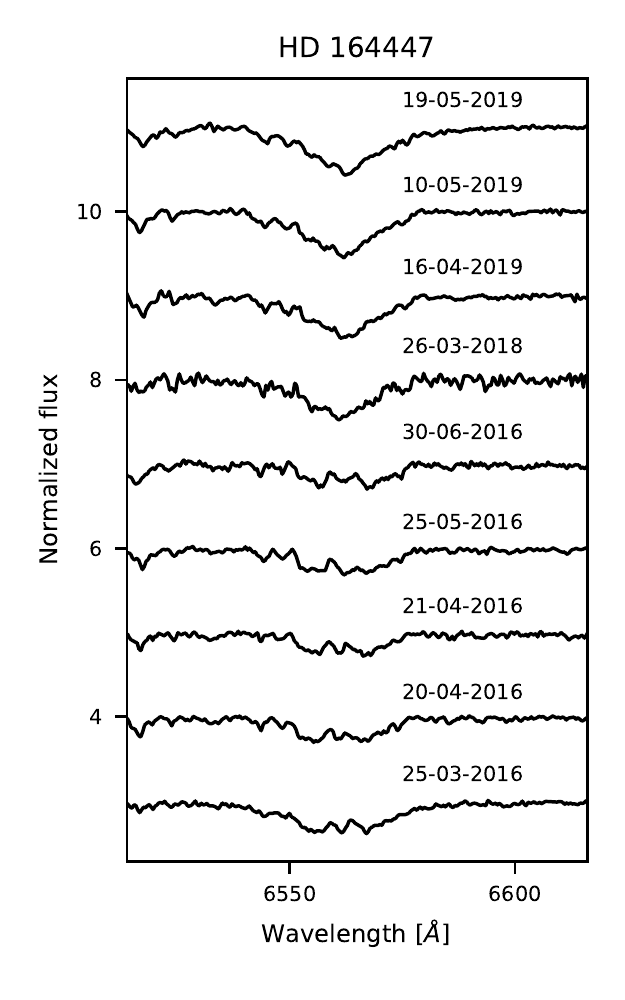}
  \end{minipage}
 \begin{minipage}[b]{0.44\textwidth}
    \includegraphics[height=78mm, width=50mm]{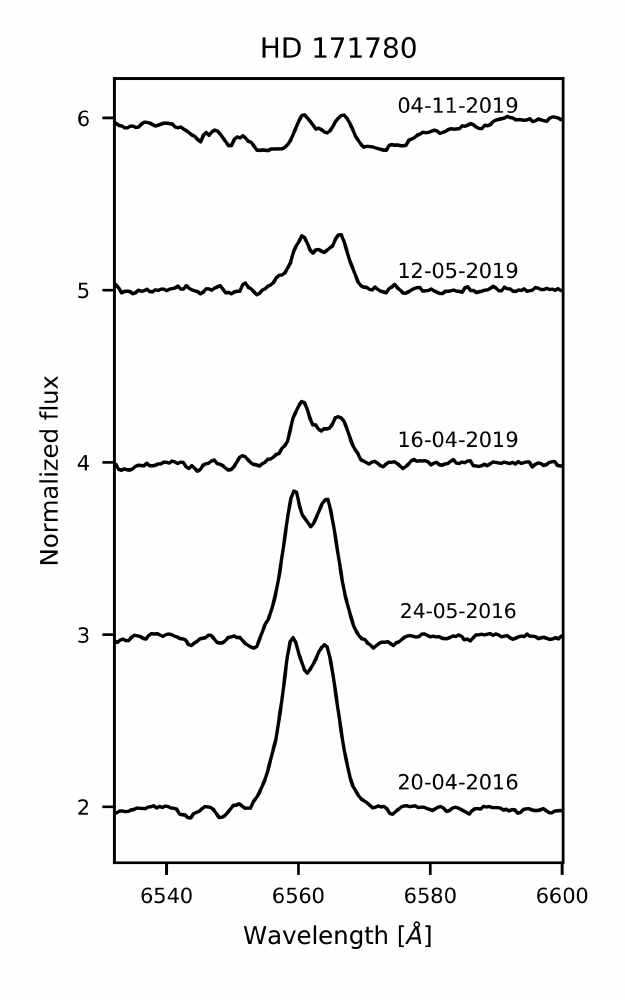}
  \end{minipage}
 \caption{H$\alpha$ line profile variation observed for HD 38708, HD 60855, HD 142926, HD 164447 and HD 171780 over a timescale of 36, 14, 24, 43 and 44 months, respectively.}
\label{fig3HD38708}
\end{figure*}

We took 8 spectra of HD 171780 through the Kavalur facility and noticed double-peak emission emission of H$\alpha$ in all occasions. The emission strength showed a continuous decrease with H$\alpha$ EW as depicted in Table \ref{table1}. This can be an indicator that this star too, like HD 164447 is losing its disc gradually. Fig. \ref{fig3HD38708} presents the H$\alpha$ line profile variation observed for HD 171780 over a timescale of 44 months.

\subsection{Comparative study with data from the BeSS database}
We looked into the BeSS database \citep{2011Neiner} to check the H$\alpha$ line profile for all our program stars. We found several spectra for each star taken by different amateur astronomers in the BeSS database. Although most of the spectra closely match with the H$\alpha$ profiles observed by us through the UAGS instrument of Kavalur, some interesting profiles have also been found in this database close to our epoch of observations.

One interesting case was noticed when we looked for the star HD 60855 in BeSS. We have already mentioned that this star showed H$\alpha$ in absorption with an EW of 4.9 \AA~when we observed it with HCT on January 11, 2008 \citep{2021Banerjee}. However, H$\alpha$ was present in double-peak emission profile in the spectrum taken by the amateur astronomer Guarro Fló on March 15, 2008 (NEWTON 254 - LHIRES-B12t - AUDINE 403) from a site in Spain. A similar profile is also seen in the next spectrum available in the BeSS database (March 17, 2009; C. Buil; C11 eShel QSI532) which was taken from Castanet, France. H$\alpha$ has remained in double-peak emission profile in all other 10 spectra since 2015 taken by different amateur astronomers. Then 2016 onward, we were able to obtain the spectra of this star using the Kavalur facility. As mentioned earlier, we also found H$\alpha$ in double-peak emission in every occasion. Similar profile was noticed by amateur astronomer Garde (RC400 Astrosib-Eshel-ATIK460EX; France) too in the 2 spectra whose date of observations (January 02 and February 17, 2017) coincide with ours.

For another star HD 23302, amateur astronomer Houpert found that emission in absorption profile for H$\alpha$ was visible since November 3, 2018 (C11-LHIRESIII194-2400t-35-QSI516s; France). Unfortunately, we could not obtain any spectrum of this star during nearby epochs. Later using the Kavalur facility, we also observed H$\alpha$ in emission in absorption profile when we were able to observe the star again on February 18, 2019. Henceforth, HD 23302 exhibited emission in absorption profile of H$\alpha$ until the last date of our observation (November 04, 2019). One more star, HD 38708 displayed H$\alpha$ in shell profile when observed by amateur astronomer Lester (September 16, 2016; 31cmDK+23um1800lpm+QSI583; Canada). We do not have the spectra of this star in nearby epochs.

\subsection{Epoch-wise variation of the H$\alpha$ Equivalent Width}
For further checking the nature of variability occurring in the discs of our program stars, we studied the H$\alpha$ EW for each of them against their corresponding dates of observation (in MJD). This exercise helped us to check the overall variation of the H$\alpha$ EW for all 9 stars as noticed throughout our observation time. For each star the variation of H$\alpha$ EW is considered based on our observation time from start to end.

\begin{figure*}
  \centering
  \begin{minipage}[b]{0.4\textwidth}
    \includegraphics[width=\linewidth,height=\textheight,keepaspectratio]{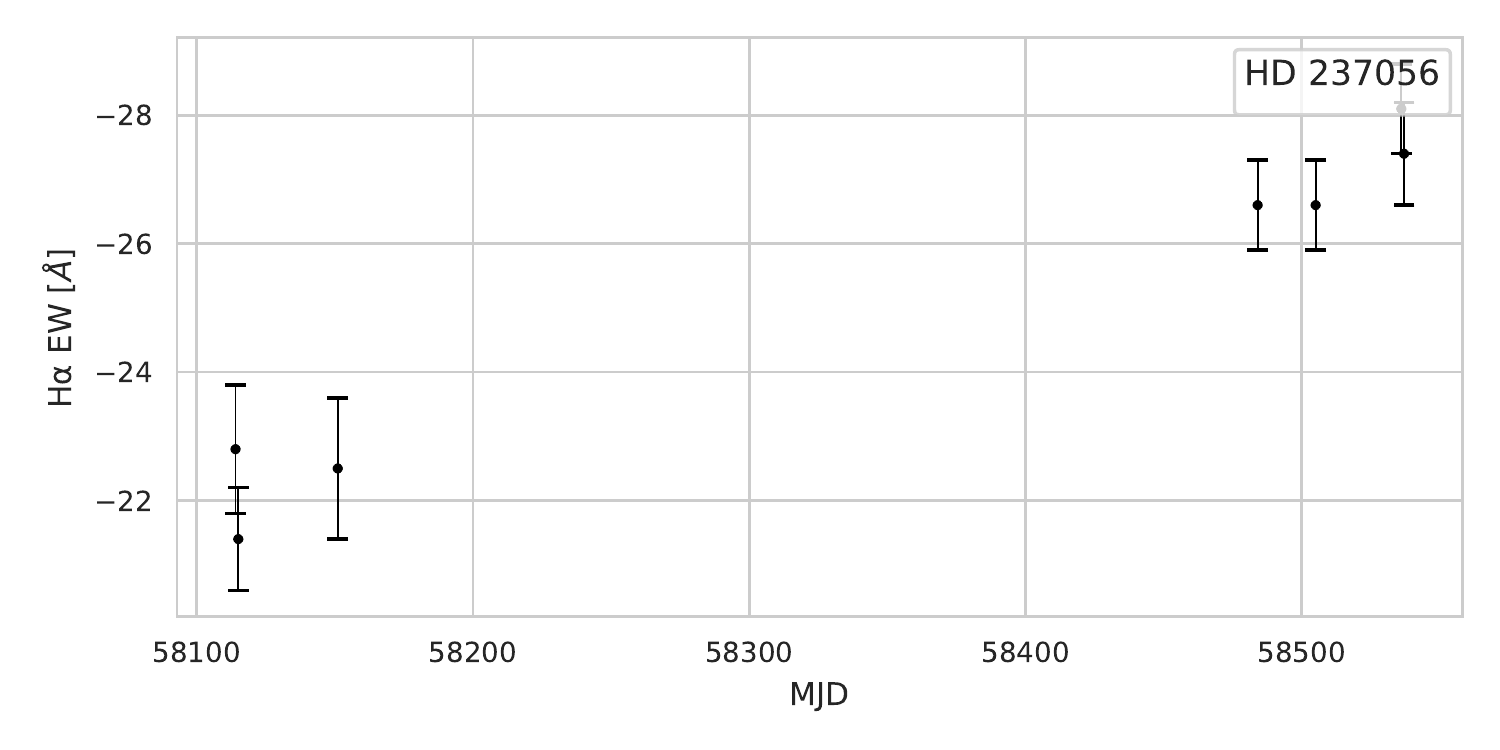}
  \end{minipage}
  \hspace{1em}
  \begin{minipage}[b]{0.4\textwidth}
    \includegraphics[width=\linewidth,height=\textheight,keepaspectratio]{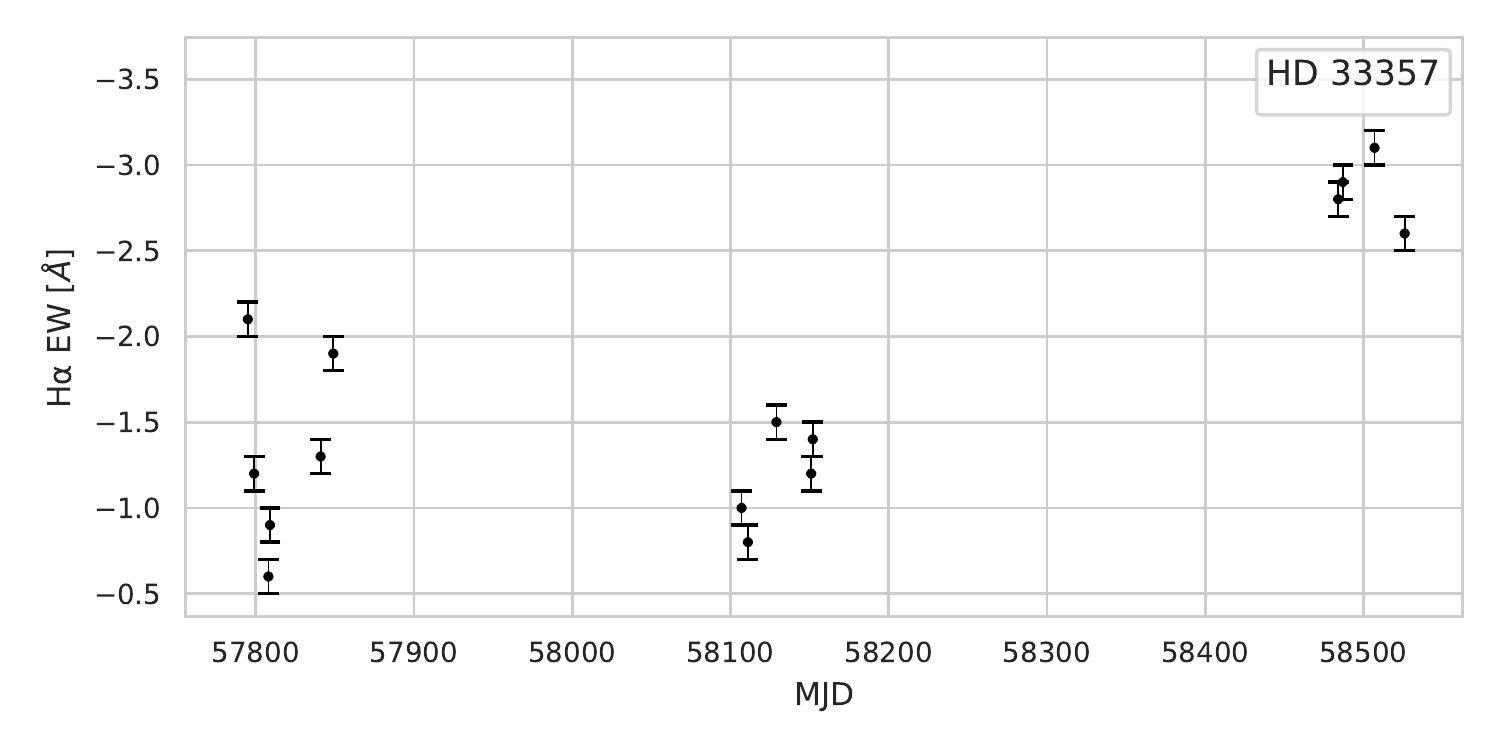}
  \end{minipage}
  \begin{minipage}[b]{0.4\textwidth}
    \includegraphics[width=\linewidth,height=\textheight,keepaspectratio]{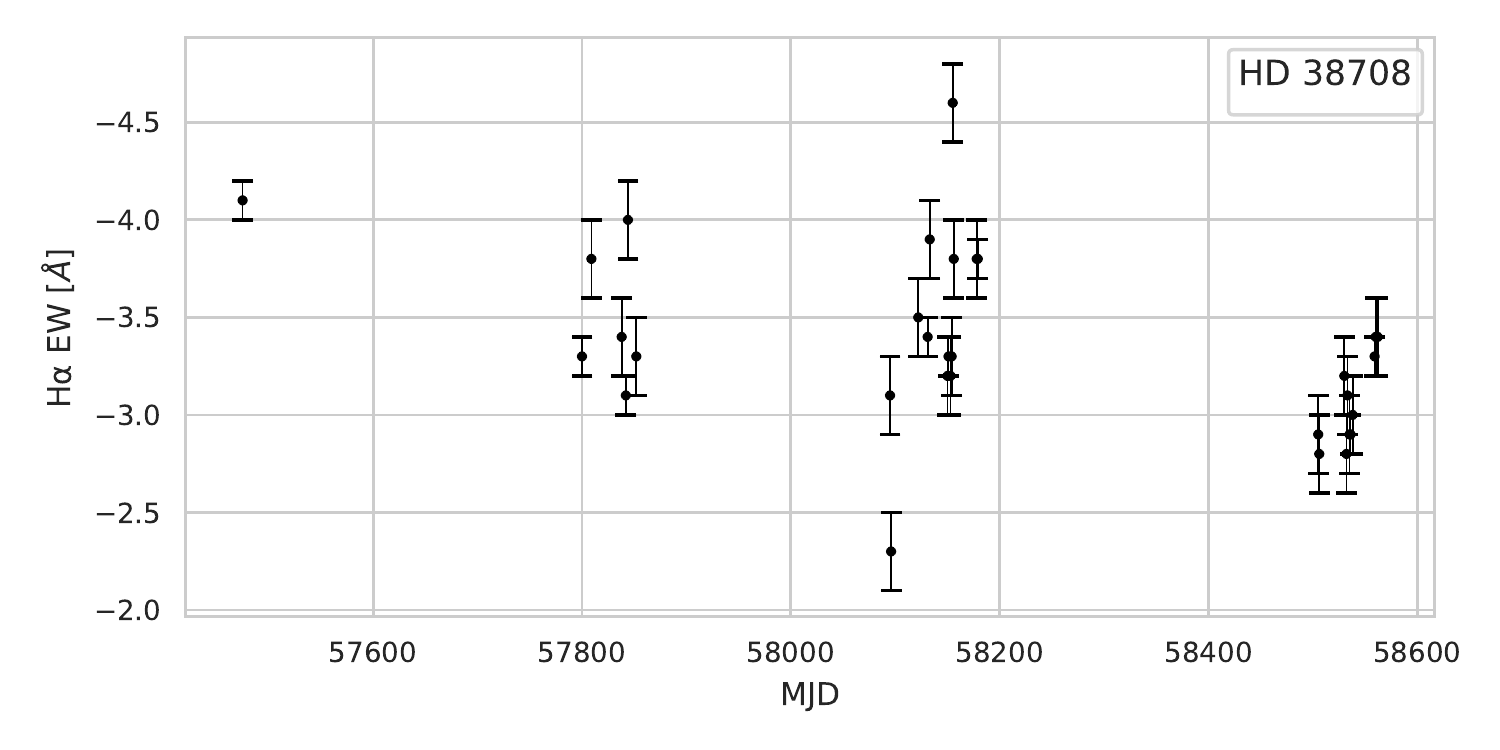}
  \end{minipage}
  \hspace{1em}
  \begin{minipage}[b]{0.4\textwidth}
    \includegraphics[width=\linewidth,height=\textheight,keepaspectratio]{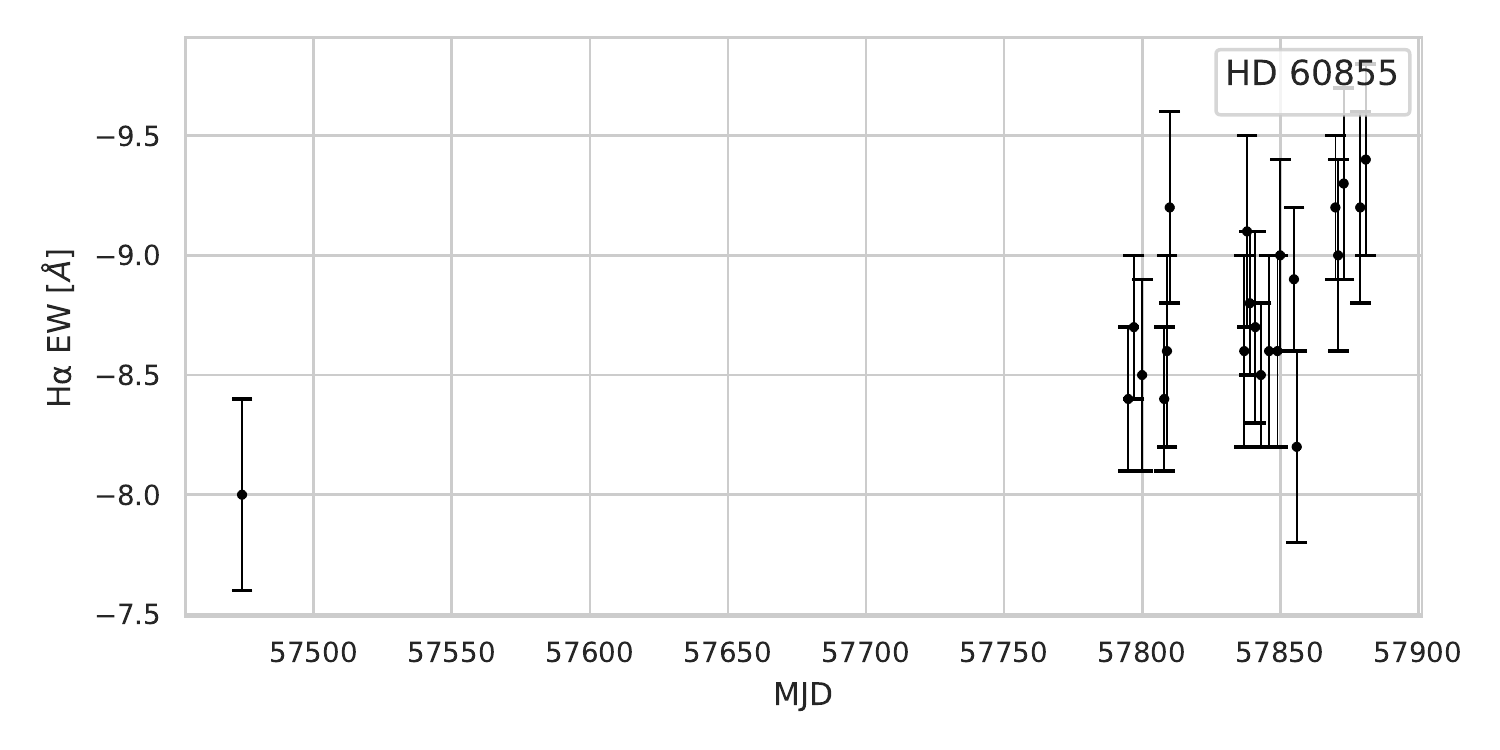}
  \end{minipage}
 \caption{Epoch-wise variation of the H$\alpha$ EW against MJD as observed for the stars HD 237056, HD 33357, HD 38708 and HD 60855. The vertical lines represent the error bar for each data point. Through visual inspection it is seen that all these 4 stars might be possessing a stable disc during our observation periods.}
\label{fig4Quadplot}
\end{figure*}

\begin{figure*}
  \centering
  \begin{minipage}[b]{0.4\textwidth}
     \includegraphics[width=\linewidth,height=\textheight,keepaspectratio]{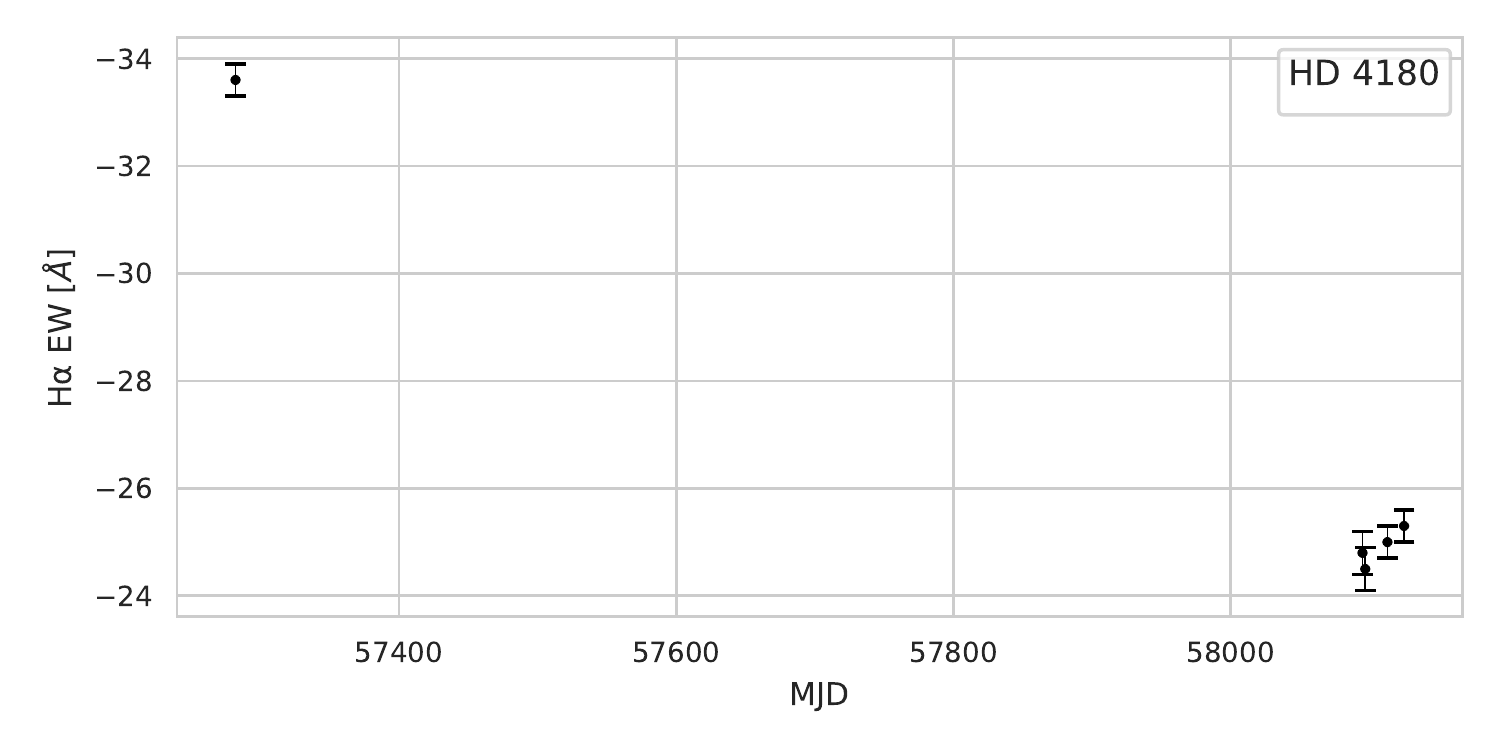}
  \end{minipage}
  \hspace{1em}
  \begin{minipage}[b]{0.4\textwidth}
    \includegraphics[width=\linewidth,height=\textheight,keepaspectratio]{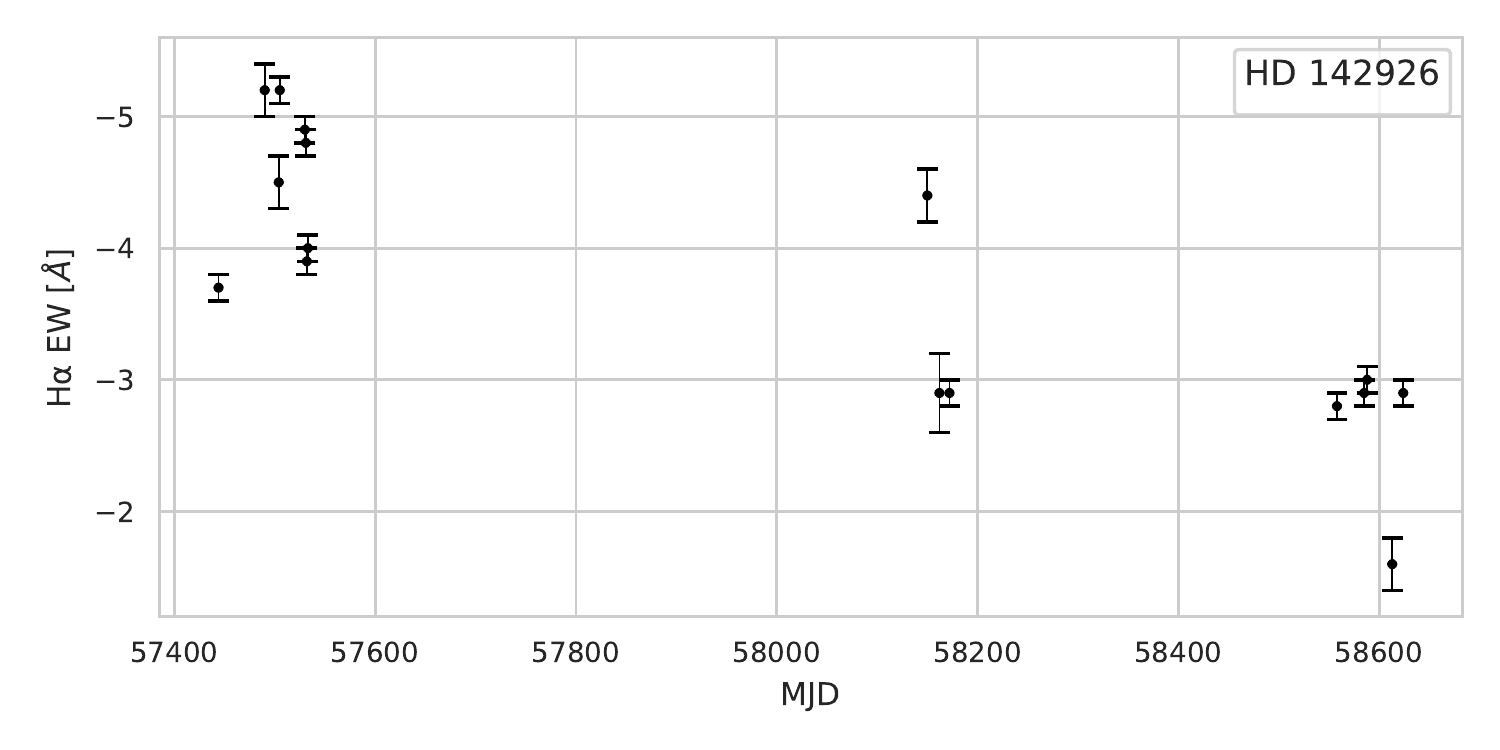}
  \end{minipage}
  \begin{minipage}[b]{0.4\textwidth}
   \includegraphics[width=\linewidth,height=\textheight,keepaspectratio]{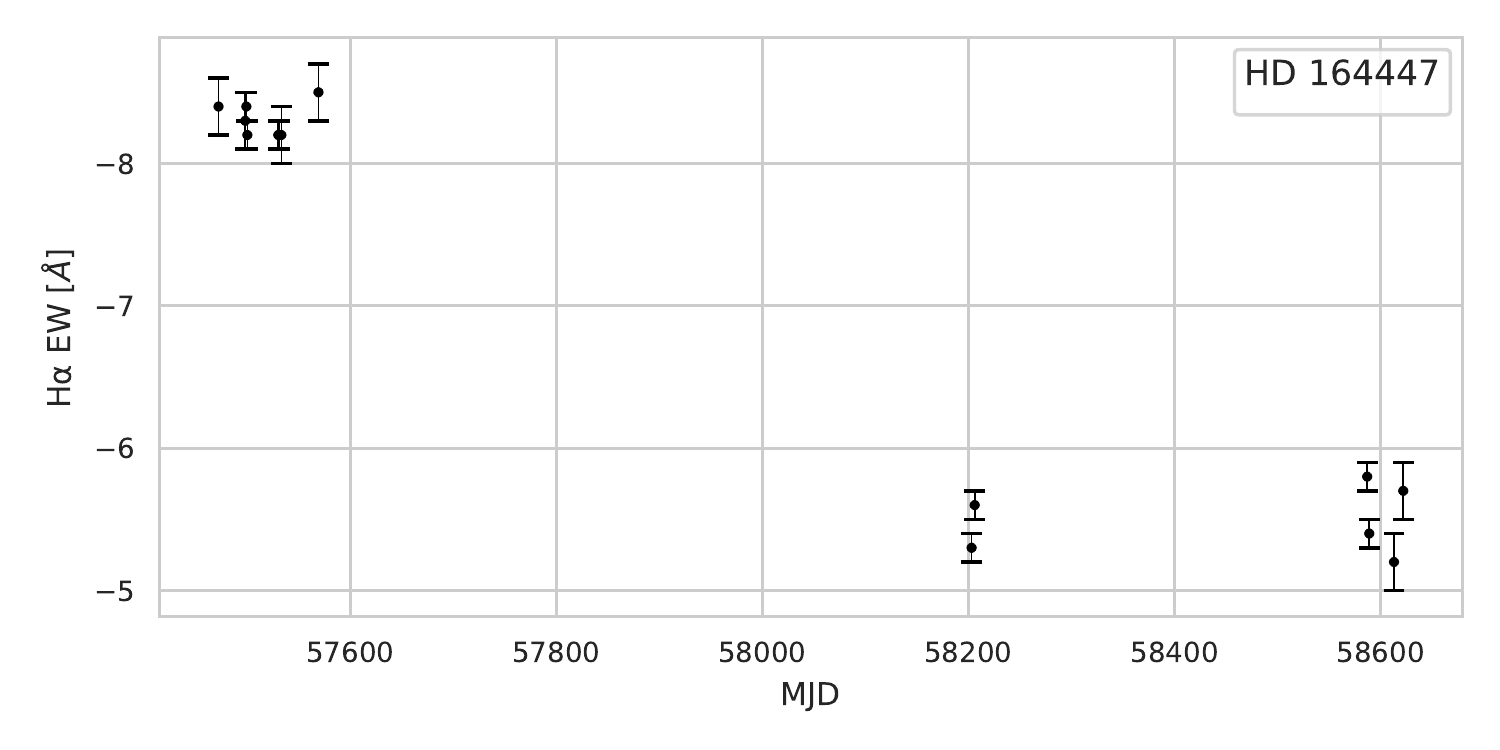}
  \end{minipage}
  \hspace{1em}
  \begin{minipage}[b]{0.4\textwidth}
    \includegraphics[width=\linewidth,height=\textheight,keepaspectratio]{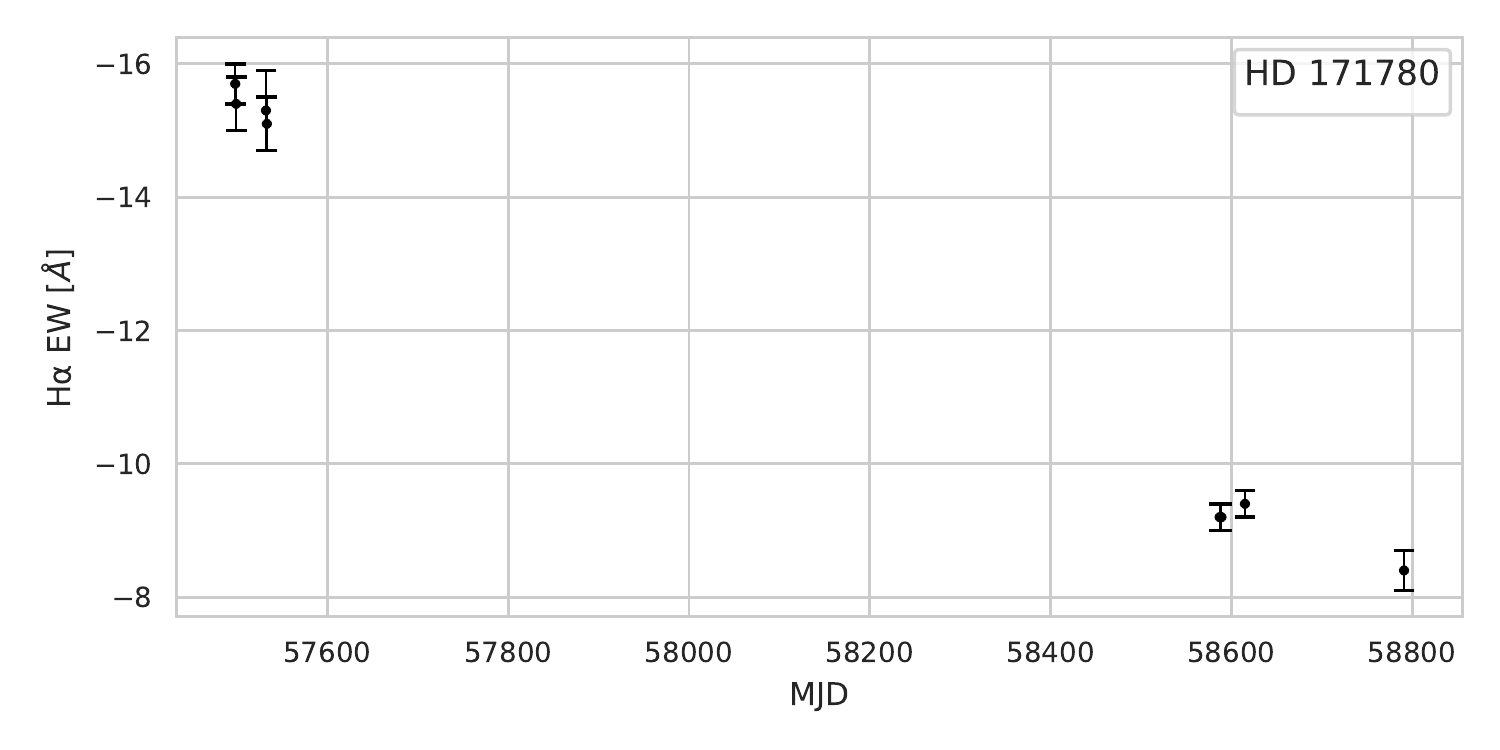}
  \end{minipage}
 \caption{Epoch-wise variation of the H$\alpha$ EW against MJD as observed for the stars HD 4180, HD 142926, HD 164447 and HD 171780. Based on our observations, it is seen that all these 4 stars exhibited an overall decrease in H$\alpha$ EW indicating that they might be passing through disc-loss episodes in the current epochs.}
\label{fig5discloss}
\end{figure*}

\begin{figure} 
\centering
  \includegraphics[width=\linewidth,height=\textheight,keepaspectratio]{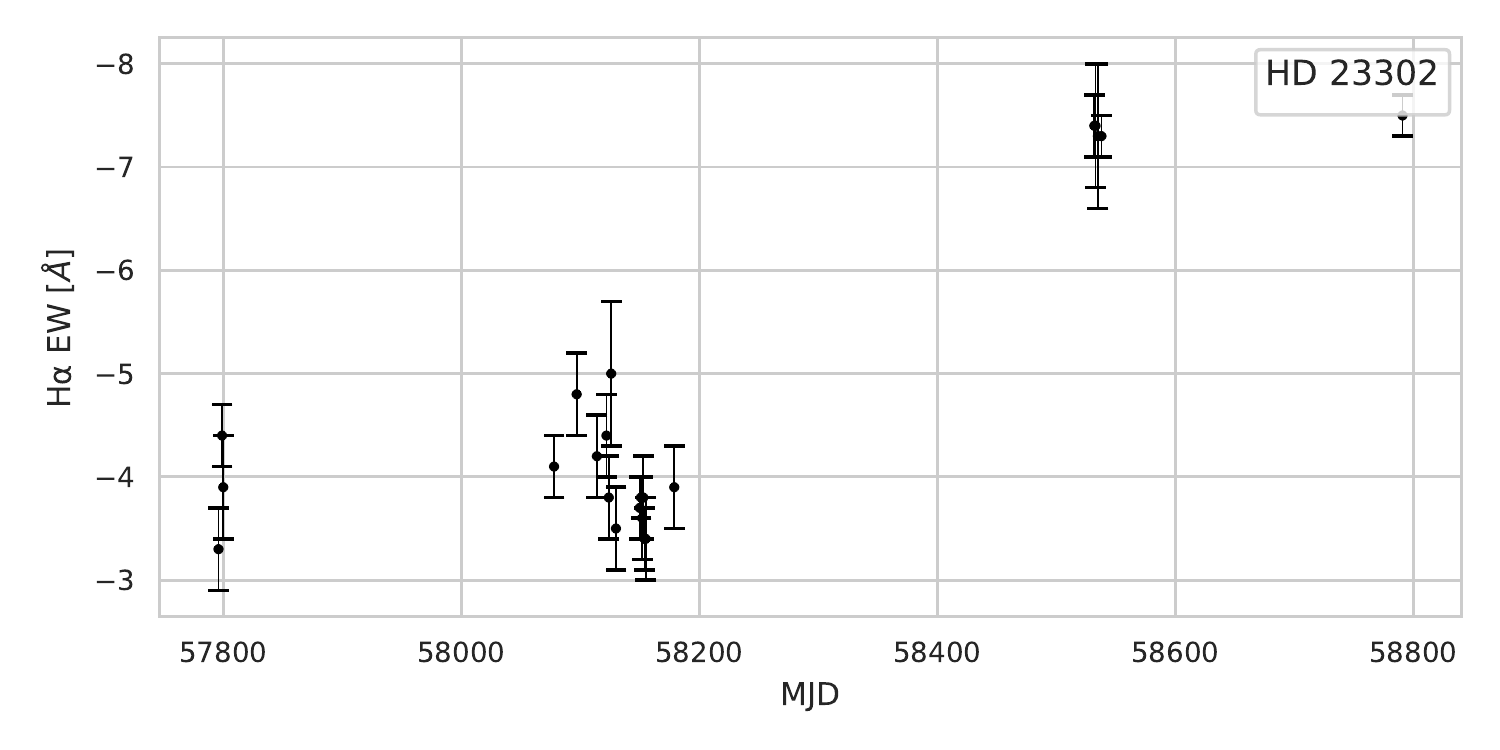}
\caption{Epoch-wise variation of the H$\alpha$ EW against MJD as observed for the stars HD 23302. It shows a trend of increasing H$\alpha$ EW since February 18, 2019 (MJD 58532). This suggests that HD 23302 might be passing through a phase of disc formation in the current epochs.}
\label{fig6forming}
\end{figure}

Figs. \ref{fig4Quadplot}, \ref{fig5discloss} and \ref{fig6forming} present the epoch-wise variation of the H$\alpha$ EW against MJD as observed for all 9 of our program stars. The vertical lines in the plots represent the error bar for each data point. We considered the error of our calculated H$\alpha$ EW values to be within 8\% in those cases where the SNR value near the H$\alpha$ line is found to be $\geq$ 90. On contrary, in cases where the SNR value is $<$ 90, the error is considered to be within 10\%.

From Fig. \ref{fig4Quadplot}, through visual inspection it is seen that for the star HD 237056, H$\alpha$ EW has increased since we started observing it. However, the overall variation of H$\alpha$ EW is moderate and remains within 20\% while comparing its multi-epoch spectra. A similar trend is exhibited by the next star HD 33357 also. In case of the rest two stars, HD 38708 and HD 60855, H$\alpha$ EW is again exhibiting moderate variation and remains within 20\%. Although HD 60855 was observed after a gap of over 300 days (duration between the first and second date of our observations as mentioned in Table \ref{table1}), the variation of H$\alpha$ EW is noticed to remain well within 20\% in all spectra obtained for this star. Our observations thus indicate that these 4 stars might be possessing a stable disc during our observation periods. 

On the contrary, the stars HD 4180, HD 142926, HD 164447 and HD 171780 exhibited an overall decrease in H$\alpha$ EW as observed by us (Fig. \ref{fig5discloss}). Hence, our observations can be suggestive of the fact that these 4 stars might be might be passing through disc-loss episodes in the current epochs. Only one star among our sample, HD 23302 shows a trend of increasing H$\alpha$ EW (Fig. \ref{fig6forming}) since February 18, 2019 (MJD 58532). This indicates that HD 23302 might be passing through a phase of disc formation in the current epochs.

So based on our observations, we classified the program stars in 3 categories, i.e., i) Group I -- those stars which possess a stable disc, ii) Group II -- stars where disc building happens during the period of observation, and iii) Group III -- stars which undergo disc dissipation, as evident from a reduction in H$\alpha$ EW during the observation period. We found that while 4 stars (HD 237056, HD 33357, HD 38708 and HD 60855) fall in Group I, HD 23302 belong to Group II and the remaining 4 stars (HD 4180, HD 142926, HD 164447 and HD 171780) fall in Group III category, respectively.

\subsection{Stars undergoing disc-loss episodes}
The present study helped us to identify the transient nature in the disc of 9 selected Galactic CBe stars. Our results indicate that 4 among these 9 stars, HD 4180, HD 142926, HD 164447 and HD 171780, are possibly losing their circumstellar disc gradually. Carefully examining Table \ref{table1}, it is noticed that HD 142926 and HD 164447 showed this trend of disc dissipation within a timescale of 37 (April 2016 -- May 2019) and 38 (March 2016 -- May 2019) months, respectively. The star HD 171780 exhibited such a trend within a timescale of 43 months (April 2016 -- November 2019), as is observed from Table \ref{table1}. Unfortunately, we could not detect any such timescale in the case of HD 4180 due to the lack of data. However, we found that the disc-loss for all these 4 stars are still continuing which is clearly visible from Table \ref{table1}. Our results thus suggest that the disc-loss episode for CBe stars can continue for a period of over 3 years.

Another star, HD 60855 showed signs of possessing a stable disc in recent epochs. However, after comparing with \cite{2021Banerjee} and the BeSS database, we were able to detect that this star has passed through a disc-less episode during January 2008. The disc was not present when we observed it in 2008 with HCT \citep{2021Banerjee}. Then during our present dates of observation, H$\alpha$ was found to exhibit a double-peak emission profile in every case with a minor variation from -8 to -9.4 \AA, which is about 20\% of the mean value. Next, from the BeSS database, we found one spectrum of this star taken on March 15, 2008 which exhibited H$\alpha$ in double-peak emission profile. This result points out that the disc formation for HD 60855 might have taken place within a timescale of only 2 months, between January -- March 2008.

\begin{figure}[h]
\centering
   \includegraphics[height=85mm, width=55mm]{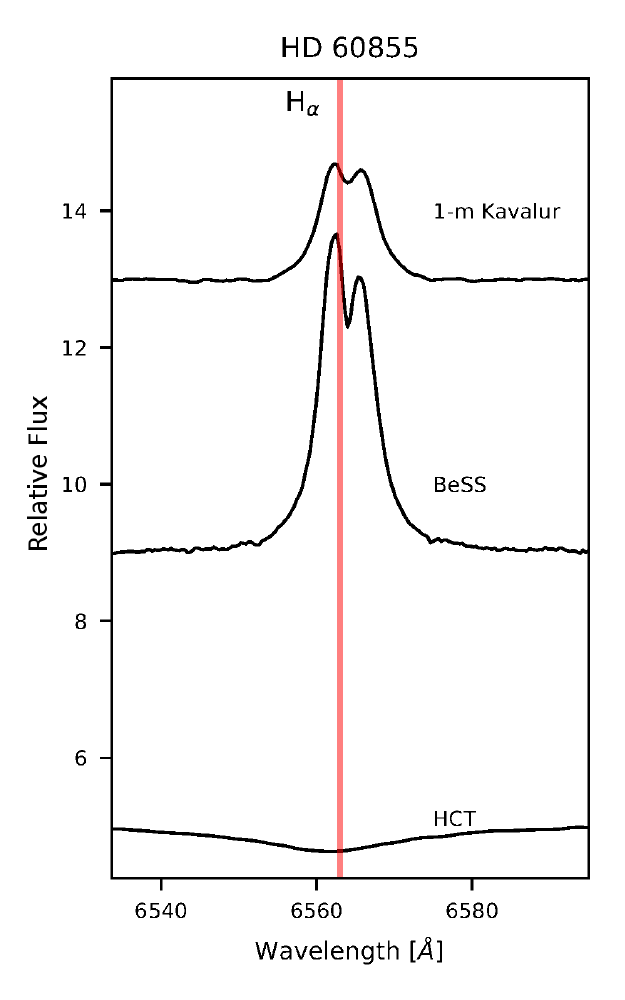}
\caption{H$\alpha$ line profile variation observed for HD 60855 by us in two occasions and the BeSS database in one case. The bottom-most spectrum shown here was taken by us using the HFOSC instrument mounted on the HCT on January 11, 2008. It showed H$\alpha$ below the continuum and we found H$\alpha$ corrected EW to be 4.9 \AA~indicating that the disc was not present during that epoch. From the BeSS database, we found another spectrum of this star (shown in this plot in the middle panel) taken by the amateur astronomer Guarro Fló (NEWTON 254 - LHIRES-B12t - AUDINE 403) on March 15, 2008 which exhibited H$\alpha$ in double-peak emission profile. Next, the topmost spectrum presented here was taken on March 27, 2016 by us with the UAGS instrument of the 1-m telescope from VBO, Kavalur. It is visible that H$\alpha$ was observed to be in double-peak emission profile.}
\label{fig5variation}
\end{figure}

Fig. \ref{fig5variation} presents the H$\alpha$ line profile variation observed for HD 60855 by us and the BeSS database. The bottom-most spectrum shown in the figure was taken by us using the HFOSC instrument mounted on the HCT on January 11, 2008. It showed H$\alpha$ below the continuum and we found H$\alpha$ corrected EW to be 4.9 \AA~indicating that the disc was not present during that epoch. From the BeSS database, we found another spectrum of this star (shown in the middle panel of the figure) taken by the amateur astronomer Guarro Fló (NEWTON 254 - LHIRES-B12t - AUDINE 403) on March 15, 2008 which exhibited H$\alpha$ in double-peak emission profile. Next, the topmost spectrum presented in the plot was taken on March 27, 2016 by us with the UAGS instrument of the 1-m telescope from VBO, Kavalur. It is visible that H$\alpha$ was observed to be in double-peak emission profile. Later on, we noticed that H$\alpha$ showed double-peak emission profile on each instance taken by the Kavalur facility. This comparative study indicates that HD 60855 has passed through a disc-loss episode during the recent past. We, therefore, suggest that the disc formation for HD 60855 has taken place within a timescale of only 2 months, within January -- March 2008.

\cite{2016Kee} performed a detailed study of line-driven ablation of optically thin discs around luminous, early-type stars. They found that for optically thin CBe star discs, such ablation can cause disc destruction within a timescale of months to years. This is in agreement with our observed results. However, further analysis using larger sample is required to provide conclusive statements regarding the disc formation and dissipation timescales in CBe stars.

\section{Conclusions}
\label{Section4}
In this paper, we studied a sample of selected 9 bright CBe stars, in the wavelength range of 6200 - 6700 \AA, which have shown H$\alpha$ line in complete absorption at least once in literature, indicating that these stars have passed through a disc-less phase at least once. The prominent results we obtained are summarized below:

\begin{itemize}
\item{We found that H$\alpha$ line is present in either single or double-peak emission in all spectra for every star. Moderate variation of H$\alpha$ EW is noticed for all the stars, its value ranging from -1.4 (for HD 33357 on March 03, 2018) to -33.6 Å (for HD 4180 on September 17, 2015). Apart from H$\alpha$, we detected absorption lines of He{\sc i} 6678 \AA~and Si{\sc ii} 6347, 6371 \AA~occasionally} for our sample stars.
\end{itemize}

\begin{itemize}
\item{Our results suggest that 4 among 9 of our program stars (HD 4180, HD 142926, HD 164447 and HD 171780) are possibly undergoing disc-loss episodes, whereas the star HD 23302 might be passing through a disc formation phase in current epochs. Another 4 stars (HD 237056, HD 33357, HD 38708 and HD 60855) have shown signs of possessing a stable disc in recent epochs.}
\end{itemize}

\begin{itemize}
\item{Based on our observations of the overall variation of the H$\alpha$ EW for all 9 stars,  we classified them in 3 categories, i.e., i) Group I -- those stars which possess a stable disc, ii) Group II -- stars where disc building happens during the period of observation, and iii) Group III -- stars which undergo disc dissipation, as evident from a reduction in H$\alpha$ EW during the observation period. We found that while 4 stars (HD 237056, HD 33357, HD 38708 and HD 60855) fall in Group I, HD 23302 belong to Group II and the remaining 4 stars (HD 4180, HD 142926, HD 164447 and HD 171780) fall in Group III category, respectively.}
\end{itemize}

\begin{itemize}
\item{Comparing with \cite{2021Banerjee} and the BeSS database, we are able to identify that one star, HD 60855 has passed through a disc-less episode during January, 2008. Our analysis indicates that the disc formation for HD 60855 has taken place within a timescale of only 2 months, between January to March 2008.}
\end{itemize}


\begin{itemize}
\item{The present study also points out that two of our sample stars, HD 33357 and HD 38708, might be weak H$\alpha$ emitters in nature having H$\alpha$ EW always $<$ -5 \AA.}
\end{itemize}

\section*{Acknowledgements}
We would like to thank the Science \& Engineering Research Board (SERB), a statutory body of the Department of Science \& Technology (DST), Government of India, for funding our research under grant number EMR/2016/006823. We thank the Center for Research, CHRIST (Deemed to be University), Bangalore, India, for funding our research which was a MRP in nature. Next, we convey our heartiest gratitude to the staff members of the Vainu Bappu Observatory (VBO), Kavalur for continuous monitoring of our program stars and obtaining their spectra with the UAGS instrument. Thanks is also deserved by databses such as the SIMBAD database and the online VizieR library service for helping us in literature survey and obtaining relevant data. This work has made use of the BeSS database, operated at LESIA, Observatoire de Meudon, France (http://basebe.obspm.fr). Hence, we thank the BeSS database too.

\bibliographystyle{apj}
\bibliography{bibtex}

\centering
\begin{table*}
\caption{List of our program CBe stars and the log of observations. Our measured and absorption corrected H$\alpha$ EW for every stars on each occasion are shown in Columns 9 and 10. The (-) sign in these columns denote emission, whereas positive value denotes absorption. The terms {\it dpe} and {\it eia} seen in some cases in Column 9 stands for ‘double-peak emission' and ‘emission in absorption', respectively. The two stars (HD 60855 and HD 171780) showing {\it dpe} profile of H$\alpha$ line on all occasions are marked with star in Column 1.}
\label{table1}
\resizebox{2.0\columnwidth}{!}{
\begin{tabular}[t]{cccccccccc}
\hline 
\multicolumn{1}{|c|}{\textbf{HD number}}&\multicolumn{1}{c|}{\textbf{Alias}}&\multicolumn{1}{c|}{\textbf{RA}}&\multicolumn{1}{c|}{\textbf{Dec}}&\multicolumn{1}{c|}{\textbf{V mag}}&\multicolumn{1}{c}{\textbf{Sp. type}}&\multicolumn{1}{|c|}{\textbf{Date of observation}}&\multicolumn{1}{c|}{\textbf{Exp. time}}&\multicolumn{1}{c|}{\textbf{H$\alpha$}}&\multicolumn{1}{c|}{\textbf{H$\alpha$}}\\
\multicolumn{1}{|c|}{\textbf{ }}&\multicolumn{1}{c|}{\textbf{}}&\multicolumn{1}{c|}{\textbf{(hh mm ss)}}&\multicolumn{1}{c|}{\textbf{(dd mm ss)}}&\multicolumn{1}{c}{}&\multicolumn{1}{|c|}{\textbf{}}&\multicolumn{1}{c|}{\textbf{dd/mm/yr}}&\multicolumn{1}{c|}{\textbf{(m)}}&\multicolumn{1}{c|}{\textbf{EW\textunderscore m (\AA)}}&\multicolumn{1}{c|}{\textbf{EW\textunderscore c (\AA)}}\\
\hline
HD 4180	&	Omi Cas	&	00 44 43.52	&	+48 17 03.71	&	4.5	&	B5IIIe	&	17/09/15	& 15		&	-27.2	&	-33.6	\\
	&		&		&		&		&		&	08/12/17	&		&	-18.4	&	-24.8	\\
	&		&		&		&		&		&	10/12/17	&		&	-18.1	&	-24.5	\\
	&		&		&		&		&		&	26/12/17	&		&	-18.6	&	-25	\\
	&		&		&		&		&		&	07/01/18	&		&	-18.9	&	-25.3	\\
HD 237056	&	BD+57 681	&	03 02 37.88	&	+57 36 46.06	&	8.9	&	B0.5Vpe	&	27/12/17	& 40		&	-18.9	&	-22.8	\\
	&		&		&		&		&		&	28/12/17	&		&	-17.5	&	-21.4	\\
	&		&		&		&		&		&	02/02/18	&		&	-18.6	&	-22.5	\\
	&		&		&		&		&		&	01/01/19	&		&	-22.7	&	-26.6	\\
	&		&		&		&		&		&	22/01/19	&		&	-22.7	&	-26.6	\\
	&		&		&		&		&		&	22/02/19	&		&	-24.2	&	-28.1	\\
	&		&		&		&		&		&	23/02/19	&		&	-23.5	&	-27.4	\\
HD 23302	&	17 Tau	&	03 44 52.54	&	+24 06 48.01	&	3.7	&	B6IIIe	&	12/02/17	& 10		&	3.6	&	-3.3	\\
	&		&		&		&		&		&	15/02/17	&		&	2.5	&	-4.4	\\
	&		&		&		&		&		&	16/02/17	&		&	3	&	-3.9	\\
	&		&		&		&		&		&	21/11/17	&		&	2.8	&	-4.1	\\
	&		&		&		&		&		&	10/12/17	&		&	2.1	&	-4.8	\\
	&		&		&		&		&		&	27/12/17	&		&	2.7	&	-4.2	\\
	&		&		&		&		&		&	04/01/18	&		&	2.5	&	-4.4	\\
	&		&		&		&		&		&	06/01/18	&		&	3.1	&	-3.8	\\
	&		&		&		&		&		&	08/01/18	&		&	1.9	&	-5	\\
	&		&		&		&		&		&	12/01/18	&		&	3.4	&	-3.5	\\
	&		&		&		&		&		&	01/02/18	&		&	3.2	&	-3.7	\\
	&		&		&		&		&		&	02/02/18	&		&	3.1	&	-3.8	\\
	&		&		&		&		&		&	03/02/18	&		&	3.3	&	-3.6	\\
	&		&		&		&		&		&	04/02/18	&		&	3.1	&	-3.8	\\
	&		&		&		&		&		&	05/02/18	&		&	3.5	&	-3.4	\\
	&		&		&		&		&		&	06/02/18	&		&	3.5	&	-3.4	\\
	&		&		&		&		&		&	02/03/18	&		&	3	&	-3.9	\\
	&		&		&		&		&		&	18/02/19	&		&	-0.5 ({\it eia, dpe})	&	-7.4	\\
	&		&		&		&		&		&	19/02/19	&		&	-0.5 ({\it eia, dpe})	&	-7.4	\\
	&		&		&		&		&		&	21/02/19	&		&	-0.4 ({\it eia, dpe})	&	-7.3	\\
	&		&		&		&		&		&	24/02/19	&		&	-0.4 ({\it eia, dpe})	&	-7.3	\\
	&		&		&		&		&		&	04/11/19	&		&	-0.5 ({\it eia, dpe})	&	-7.5	\\
HD 33357	&	SX Aur	&	05 11 42.93	&	+42 09 55.28	&	8.6	&	B1Vne	&	11/02/17	& 40		&	1.8	&	-2.1	\\
	&		&		&		&		&		&	15/02/17	&		&	2.7	&	-1.2	\\
	&		&		&		&		&		&	24/02/17	&		&	3.3	&	-0.6	\\
	&		&		&		&		&		&	25/02/17	&		&	3	&	-0.9	\\
	&		&		&		&		&		&	29/03/17	&		&	2.6	&	-1.3	\\
	&		&		&		&		&		&	06/04/17	&		&	2	&	-1.9	\\
	&		&		&		&		&		&	20/12/17	&		&	2.9	&	-1	\\
	&		&		&		&		&		&	24/12/17	&		&	3.1	&	-0.8	\\
	&		&		&		&		&		&	11/01/18	&		&	2.4	&	-1.5	\\
		&		&		&		&		&		&	02/02/18	&		&	2.7	&	-1.2	\\
	&		&		&		&		&		&	03/03/18	&		&	2.5	&	-1.4	\\
	&		&		&		&		&		&	01/01/19	&		&	1.1	&	-2.8	\\
	&		&		&		&		&		&	04/01/19	&		&	1	&	-2.9	\\
	&		&		&		&		&		&	24/01/19	&		&	0.8	&	-3.1	\\
	&		&		&		&		&		&	12/02/19	&		&	1.3	&	-2.6	\\
HD 38708	&	V438 Aur	&	05 48 53.65	&	+29 08 10.02	&	8.2	&	B3/4Vn(e)	&	28/03/16	& 40		&	1.5	&	-4.1	\\
	&		&		&		&		&		&	16/02/17	&		&	2.3	&	-3.3	\\
	&		&		&		&		&		&	25/02/17	&		&	1.8	&	-3.8	\\
	&		&		&		&		&		&	26/03/17	&		&	2.2	&	-3.4	\\
	&		&		&		&		&		&	30/03/17	&		&	2.5	&	-3.1	\\
	&		&		&		&		&		&	01/04/17	&		&	1.6	&	-4	\\
	&		&		&		&		&		&	09/04/17	&		&	2.3	&	-3.3	\\
	&		&		&		&		&		&	08/12/17	&		&	2.5	&	-3.1	\\
	&		&		&		&		&		&	09/12/17	&		&	3.3	&	-2.3	\\
		&		&		&		&		&		&	04/01/18	&		&	2.1	&	-3.5	\\
	&		&		&		&		&		&	13/01/18	&		&	2.2	&	-3.4	\\
	&		&		&		&		&		&	15/01/18	&		&	1.9	&	-3.9	\\
	&		&		&		&		&		&	01/02/18	&		&	2.4	&	-3.2	\\
	&		&		&		&		&		&	02/02/18	&		&	2.3	&	-3.3	\\
	&		&		&		&		&		&	04/02/18	&		&	2.4	&	-3.2	\\
		&		&		&		&		&		&	05/02/18	&		&	2.3	&	-3.3	\\
	&		&		&		&		&		&	06/02/18	&		&	2	&	-4.6	\\
	&		&		&		&		&		&	08/02/18	&		&	1.8	&	-3.8	\\
\hline
\end{tabular}}
\end{table*}

\begin{table*}
\centering
\label{tab:table1}
\resizebox{2.0\columnwidth}{!}{
\begin{tabular}[t]{cccccccccc}
\hline 
\multicolumn{1}{|c|}{\textbf{HD number}}&\multicolumn{1}{c|}{\textbf{Alias}}&\multicolumn{1}{c|}{\textbf{RA}}&\multicolumn{1}{c|}{\textbf{Dec}}&\multicolumn{1}{c|}{\textbf{V mag}}&\multicolumn{1}{c}{\textbf{Sp. type}}&\multicolumn{1}{|c|}{\textbf{Date of observation}}&\multicolumn{1}{c|}{\textbf{Exp. time}}&\multicolumn{1}{c|}{\textbf{H$\alpha$}}&\multicolumn{1}{c|}{\textbf{H$\alpha$}}\\
\multicolumn{1}{|c|}{\textbf{ }}&\multicolumn{1}{c|}{\textbf{}}&\multicolumn{1}{c|}{\textbf{(hh mm ss)}}&\multicolumn{1}{c|}{\textbf{(dd mm ss)}}&\multicolumn{1}{c}{}&\multicolumn{1}{|c|}{\textbf{}}&\multicolumn{1}{c|}{\textbf{dd/mm/yr}}&\multicolumn{1}{c|}{\textbf{(m)}}&\multicolumn{1}{c|}{\textbf{EW\textunderscore m (\AA)}}&\multicolumn{1}{c|}{\textbf{EW\textunderscore c (\AA)}}\\
\hline
	&		&		&		&		&		&	01/03/18	&		&	1.8	&	-3.8	\\
	&		&		&		&		&		&	02/03/18	&		&	1.8	&	-3.8	\\
	&		&		&		&		&		&	22/01/19	&		&	2.7	&	-2.9	\\
	&		&		&		&		&		&	23/01/19	&		&	2.8	&	-2.8	\\
	&		&		&		&		&		&	16/02/19	&		&	2.4	&	-3.2	\\
	&		&		&		&		&		&	18/02/19	&		&	2.8	&	-2.8	\\
	&		&		&		&		&		&	19/02/19	&		&	2.5	&	-3.1	\\
	&		&		&		&		&		&	21/02/19	&		&	2.7	&	-2.9	\\
	&		&		&		&		&		&	22/02/19	&		&	2.7	&	-2.9	\\
	&		&		&		&		&		&	24/02/19	&		&	2.6	&	-3	\\
	&		&		&		&		&		&	17/03/19	&		&	2.3	&	-3.3	\\
	&		&		&		&		&		&	18/03/19	&		&	2.2	&	-3.4	\\
	&		&		&		&		&		&	22/03/19	&		&	2.2	&	-3.4	\\
{HD 60855}$^\star$	&	V378 Pup	&	07 36 03.89	&	-14 29 33.9	&	5.7	&	B2Ve	&	27/03/16	& 20		&	-3.3	&	-8	\\
	&		&		&		&		&		&	11/02/17	&		&	-3.7	&	-8.4	\\
	&		&		&		&		&		&	13/02/17	&		&	-4	&	-8.7	\\
	&		&		&		&		&		&	16/02/17	&		&	-3.8	&	-8.5	\\
	&		&		&		&		&		&	24/02/17	&		&	-3.7	&	-8.4	\\
	&		&		&		&		&		&	25/02/17	&		&	-3.9	&	-8.6	\\
	&		&		&		&		&		&	26/02/17	&		&	-4.5	&	-9.2	\\
	&		&		&		&		&		&	25/03/17	&		&	-3.9	&	-8.6	\\
	&		&		&		&		&		&	26/03/17	&		&	-4.4	&	-9.1	\\
	&		&		&		&		&		&	27/03/17	&		&	-4.1	&	-8.8	\\
	&		&		&		&		&		&	29/03/17	&		&	-4	&	-8.7	\\
	&		&		&		&		&		&	31/03/17	&		&	-3.8	&	-8.5	\\
	&		&		&		&		&		&	03/04/17	&		&	-3.9	&	-8.6	\\
	&		&		&		&		&		&	06/04/17	&		&	-3.9	&	-8.6	\\
	&		&		&		&		&		&	07/04/17	&		&	-4.3	&	-9	\\
	&		&		&		&		&		&	12/04/17	&		&	-4.2	&	-8.9	\\
	&		&		&		&		&		&	13/04/17	&		&	-3.5	&	-8.2	\\
	&		&		&		&		&		&	27/04/17	&		&	-4.5	&	-9.2	\\
	&		&		&		&		&		&	28/04/17	&		&	-4.3	&	-9	\\
	&		&		&		&		&		&	30/04/17	&		&	-4.6	&	-9.3	\\
	&		&		&		&		&		&	06/05/17	&		&	-4.5	&	-9.2	\\
	&		&		&		&		&		&	08/05/17	&		&	-4.7	&	-9.4	\\
HD 142926	&	4 Her	&	15 55 30.59	&	+42 33 58.29	&	5.8	&	B7IV/V -- B9e	&	26/02/16	& 20		&	3.8	&	-3.7	\\
	&		&		&		&		&		&	12/04/16	&		&	2.3	&	-5.2	\\
	&		&		&		&		&		&	26/04/16	&		&	3	&	-4.5	\\
	&		&		&		&		&		&	27/04/16	&		&	2.3	&	-5.2	\\
	&		&		&		&		&		&	22/05/16	&		&	2.6	&	-4.9	\\
	&		&		&		&		&		&	23/05/16	&	& 2.7	& -4.8 \\
	&		&		&		&		&		&	24/05/16	&		&	3.6	&	-3.9	\\
	&		&		&		&		&		&	25/05/16	&		&	3.5	&	-4	\\
	&		&		&		&		&		&	01/02/18	&		&	3.1	&	-4.4	\\
	&		&		&		&		&		&	13/02/19	&		&	4.6	&	-2.9	\\
	&		&		&		&		&		&	23/02/19	&		&	4.6	&	-2.9	\\
	&		&		&		&		&		&	16/03/19	&		&	4.7	&	-2.8	\\
	&		&		&		&		&		&	12/04/19	&		&	4.6	&	-2.9	\\
	&		&		&		&		&		&	15/04/19	&		&	4.5	&	-3	\\
	&		&		&		&		&		&	10/05/19	&		&	5.9	&	-1.6	\\
	&		&		&		&		&		&	21/05/19	&		&	4.6	&	-2.9	\\
HD 164447	&	V974 Her	&	18 00 27.66	&	+19 30 20.83	&	6.4	&	B8Vn	&	25/03/16	& 25		&	-0.5 ({\it eia, dpe})	&	-8.4	\\
	&		&		&		&		&		&	20/04/16	&		&	-0.4 ({\it eia, dpe})	&	-8.3	\\
	&		&		&		&		&		&	21/04/16	&		&	-0.4 ({\it eia, dpe})	&	-8.3	\\
	&		&		&		&		&		&	22/04/16	&		&	-0.3 ({\it eia, dpe})	&	-8.2	\\
	&		&		&		&		&		&	22/05/16	&		&	-0.3 ({\it eia, dpe})	&	-8.2	\\
	&		&		&		&		&		&	23/05/16	&		&	-0.3 ({\it eia, dpe})	&	-8.2	\\
	&		&		&		&		&		&	25/05/16	&		&	-0.3 ({\it eia, dpe})	&	-8.2	\\
	&		&		&		&		&		&	30/06/16	&		&	-0.2 ({\it eia, dpe})	&	-8.1	\\
	&		&		&		&		&		&	26/03/18	&		&	2.6	&	-5.3	\\
	&		&		&		&		&		&	29/03/18	&		&	2.3	&	-5.6	\\
	&		&		&		&		&		&	14/04/19	&		&	2.1	&	-5.8	\\
	&		&		&		&		&		&	16/04/19	&		&	2.5	&	-5.4	\\
	&		&		&		&		&		&	10/05/19	&		&	2.7	&	-5.2	\\
	&		&		&		&		&		&	19/05/19	&		&	2.2	&	-5.7	\\
{HD 171780}$^\star$	&	HR 6984	&	18 35 13.51	&	+34 27 28.88	&	6.1	&	B5Ve	&	20/04/16	& 25		&	-9.3	&	-15.7	\\
	&		&		&		&		&		&	21/04/16	&		&	-9	&	-15.4	\\
	&		&		&		&		&		&	24/05/16	&		&	-8.9	&	-15.3	\\
	&		&		&		&		&		&	25/05/16	&		&	-8.7	&	-15.1	\\
	&		&		&		&		&		&	14/04/19	&		&	-2.8	&	-9.2	\\
	&		&		&		&		&		&	16/04/19	&		&	-2.8	&	-9.4	\\
	&		&		&		&		&		&	12/05/19	&		&	-3	&	-9.4	\\
	&		&		&		&		&		&	04/11/19	&		&	-2 ({\it eia})	&	-8.4	\\
\hline
\end{tabular}}
\end{table*}

\centering
\begin{table*}
\caption{Details of spectral lines observed in our program CBe stars. Column 3 presents the mean H$\alpha$ EW for every star along with associated errors. Columns 5, 6 and 7 show the presence of He{\sc i} 6678 and He{\sc i} 6347, 6371 \AA~lines on every date of our observation. Here, the symbol ‘a’ represents absorption profile, respectively.}
\label{table2}
\resizebox{2.0\columnwidth}{!}{
\begin{tabular}[t]{ccccccc}
\hline 
\multicolumn{1}{|c|}{\textbf{Star name}}&\multicolumn{1}{c|}{\textbf{Sp. type}}&\multicolumn{1}{|c|}{\textbf{Mean}}&\multicolumn{1}{c|}{\textbf{Date of observation}}&\multicolumn{1}{c|}{\textbf{He{\sc i}}}&\multicolumn{1}{c|}{\textbf{Si{\sc ii}}}&\multicolumn{1}{c|}{\textbf{Si{\sc ii}}}\\
\multicolumn{1}{|c|}{\textbf{}}&\multicolumn{1}{c|}{\textbf{}}&\multicolumn{1}{c|}{\textbf{H$\alpha$ EW (\AA)}}&\multicolumn{1}{c|}{\textbf{dd/mm/yr}}&\multicolumn{1}{c|}{\textbf{6678 \AA}}&\multicolumn{1}{c|}{\textbf{6347 \AA}}&\multicolumn{1}{c|}{\textbf{6371 \AA}}\\
\hline
HD 4180	&	B5IIIe	& -26.6 ${\pm0.4}$		&	17/09/15	&	$\times$	&	$\times$	&	$\times$	\\
	&		&		&	8/12/17	&	$\times$	&	$\times$	&	$\times$	\\
	&		&		&	10/12/17	&	$\times$	&	$\times$	&	$\times$	\\
	&		&		&	26/12/17	&	$\times$	&	$\times$	&	$\times$	\\
	&		&		&	7/01/18	&	$\times$	&	$\times$	&	$\times$	\\
HD 237056	&	B0.5Vpe	& -25.0 ${\pm0.7}$		&	27/12/17	&	$\times$	&	$\times$	&	$\times$	\\
	&		&		&	28/12/17	&	$\times$	&	$\times$	&	$\times$	\\
	&		&		&	2/02/18	&	$\times$	&	$\times$	&	$\times$	\\
	&		&		&	1/01/19	&	$\times$	&	$\times$	&	$\times$	\\
	&		&		&	22/01/19	&	$\times$	&	$\times$	&	$\times$	\\
	&		&		&	22/02/19	&	a	&	$\times$	&	$\times$	\\
	&		&		&	23/02/19	&	a	&	$\times$	&	$\times$	\\
HD 23302	&	B6IIIe	& -4.7 ${\pm0.3}$		&	24/10/15	&	$\times$	&	a	&	a	\\
	&		&		&	12/02/17	&	$\times$	&	a	&	a	\\
	&		&		&	15/02/17	&	$\times$	&	a	&	a	\\
	&		&		&	16/02/17	&	$\times$	&	a	&	a	\\
	&		&		&	21/11/17	&	$\times$	&	a	&	a	\\
	&		&		&	10/12/17	&	$\times$	&	a	&	a	\\
	&		&		&	27/12/17	&	$\times$	&	a	&	a	\\
	&		&		&	4/01/18	&	$\times$	&	a	&	a	\\
	&		&		&	6/01/18	&	$\times$	&	a	&	a	\\
	&		&		&	8/01/18	&	$\times$	&	a	&	a	\\
	&		&		&	12/01/18	&	$\times$	&	a	&	a	\\
	&		&		&	1/02/18	&	$\times$	&	a	&	a	\\
	&		&		&	2/02/18	&	$\times$	&	a	&	a	\\
	&		&		&	3/02/18	&	$\times$	&	a	&	a	\\
	&		&		&	4/02/18	&	$\times$	&	a	&	a	\\
	&		&		&	5/02/18	&	$\times$	&	a	&	a	\\
	&		&		&	6/02/18	&	$\times$	&	a	&	a	\\
	&		&		&	2/03/18	&	$\times$	&	a	&	a	\\
	&		&		&	18/02/19	&	$\times$	&	a	&	a	\\
	&		&		&	19/02/19	&	$\times$	&	a	&	a	\\
	&		&		&	21/02/19	&	$\times$	&	a	&	a	\\
	&		&		&	24/02/19	&	$\times$	&	a	&	a	\\
	&		&		&	4 Nov 19	&	$\times$	&	a	&	a	\\
HD 33357	&	B1Vne	&	-1.7 ${\pm0.1}$	&	11/02/17	&	$\times$	&	$\times$	&	$\times$	\\
	&		&		&	15/02/17	&	$\times$	&	$\times$	&	$\times$	\\
	&		&		&	24/02/17	&	$\times$	&	$\times$	&	$\times$	\\
	&		&		&	25/02/17	&	$\times$	&	$\times$	&	$\times$	\\
	&		&		&	29/03/17	&	a	&	$\times$	&	$\times$	\\
	&		&		&	6/04/17	&	a	&	$\times$	&	$\times$	\\
	&		&		&	20/12/17	&	a	&	$\times$	&	$\times$	\\
		&		&		&	24/12/17	&	a	&	$\times$	&	$\times$	\\
\hline
\end{tabular}}
\end{table*}

\begin{table*}
\centering
\label{tab:table2}
\resizebox{2.0\columnwidth}{!}{
\begin{tabular}[t]{ccccccc}
\hline 
\multicolumn{1}{|c|}{\textbf{Star name}}&\multicolumn{1}{c|}{\textbf{Sp. type}}&\multicolumn{1}{|c|}{\textbf{Mean}}&\multicolumn{1}{c|}{\textbf{Date of observation}}&\multicolumn{1}{c|}{\textbf{He{\sc i}}}&\multicolumn{1}{c|}{\textbf{Si{\sc ii}}}&\multicolumn{1}{c|}{\textbf{Si{\sc ii}}}\\
\multicolumn{1}{|c|}{\textbf{}}&\multicolumn{1}{c|}{\textbf{}}&\multicolumn{1}{c|}{\textbf{H$\alpha$ EW (\AA)}}&\multicolumn{1}{c|}{\textbf{dd/mm/yr}}&\multicolumn{1}{c|}{\textbf{6678 \AA}}&\multicolumn{1}{c|}{\textbf{6347 \AA}}&\multicolumn{1}{c|}{\textbf{6371 \AA}}\\
\hline
	&		&		&	11/01/18	&	a	&	$\times$	&	$\times$	\\
	&		&		&	2/02/18	&	a	&	$\times$	&	$\times$	\\
	&		&		&	3/03/18	&	a	&	$\times$	&	$\times$	\\
	&		&		&	1/01/19	&	a	&	$\times$	&	$\times$	\\
	&		&		&	4/01/19	&	a	&	$\times$	&	$\times$	\\
	&		&		&	24/01/19	&	a	&	$\times$	&	$\times$	\\
	&		&		&	12/02/19	&	a	&	$\times$	&	$\times$	\\
HD 38708	&	B3/4Vn(e)	& -3.3 ${\pm0.2}$		&	28/03/16	&	a	&	$\times$	&	$\times$	\\
	&		&		&	16/02/17	&	a	&	$\times$	&	$\times$	\\
	&		&		&	25/02/17	&	a	&	$\times$	&	$\times$	\\
	&		&		&	26/03/17	&	a	&	$\times$	&	$\times$	\\
	&		&		&	30/03/17	&	a	&	$\times$	&	$\times$	\\
	&		&		&	01/04/17	&	a	&	$\times$	&	$\times$	\\
	&		&		&	09/04/17	&	a	&	$\times$	&	$\times$	\\
	&		&		&	08/12/17	&	a	&	$\times$	&	$\times$	\\
	&		&		&	09/12/17	&	a	&	$\times$	&	$\times$	\\
	&		&		&	04/01/18	&	a	&	$\times$	&	$\times$	\\
	&		&		&	13/01/18	&	a	&	$\times$	&	$\times$	\\
	&		&		&	15/01/18	&	a	&	$\times$	&	$\times$	\\
	&		&		&	01/02/18	&	a	&	$\times$	&	$\times$	\\
	&		&		&	02/02/18	&	a	&	$\times$	&	$\times$	\\
	&		&		&	03/02/18	&	a	&	$\times$	&	$\times$	\\
	&		&		&	04/02/18	&	a	&	$\times$	&	$\times$	\\
	&		&		&	05/02/18	&	a	&	$\times$	&	$\times$	\\
	&		&		&	06/02/18	&	a	&	$\times$	&	$\times$	\\
	&		&		&	08/02/18	&	a	&	$\times$	&	$\times$	\\
	&		&		&	01/03/18	&	a	&	$\times$	&	$\times$	\\
	&		&		&	02/03/18	&	a	&	$\times$	&	$\times$	\\
	&		&		&	22/01/19	&	a	&	$\times$	&	$\times$	\\
	&		&		&	23/01/19	&	a	&	$\times$	&	$\times$	\\
	&		&		&	16/02/19	&	a	&	$\times$	&	$\times$	\\
	&		&		&	18/02/19	&	a	&	$\times$	&	$\times$	\\
	&		&		&	19/02/19	&	a	&	$\times$	&	$\times$	\\
	&		&		&	21/02/19	&	a	&	$\times$	&	$\times$	\\
	&		&		&	22/02/19	&	a	&	$\times$	&	$\times$	\\
	&		&		&	24/02/19	&	a	&	$\times$	&	$\times$	\\
	&		&		&	17/03/19	&	a	&	$\times$	&	$\times$	\\
	&		&		&	18/03/19	&	a	&	$\times$	&	$\times$	\\
	&		&		&	22/03/19	&	a	&	$\times$	&	$\times$	\\
HD 60855	&	B2Ve	& -8.7 ${\pm0.4}$		&	27/03/16	&	$\times$	&	$\times$	&	$\times$	\\
	&		&		&	11/02/17	&	a	&	$\times$	&	$\times$	\\
	&		&		&	13/02/17	&	a	&	$\times$	&	$\times$	\\
	&		&		&	16/02/17	&	a	&	$\times$	&	$\times$	\\
	&		&		&	24/02/17	&	a	&	$\times$	&	$\times$	\\
	&		&		&	25/02/17	&	a	&	$\times$	&	$\times$	\\
	&		&		&	26/02/17	&	a	&	$\times$	&	$\times$	\\
	&		&		&	25/03/17	&	a	&	$\times$	&	$\times$	\\
	&		&		&	26/03/17	&	$\times$	&	$\times$	&	$\times$	\\
\hline
\end{tabular}}
\end{table*}

\centering
\begin{table*}
\label{tab:table2}
\resizebox{2.0\columnwidth}{!}{
\begin{tabular}[t]{ccccccc}
\hline 
\multicolumn{1}{|c|}{\textbf{Star name}}&\multicolumn{1}{c|}{\textbf{Sp. type}}&\multicolumn{1}{|c|}{\textbf{Mean}}&\multicolumn{1}{c|}{\textbf{Date of observation}}&\multicolumn{1}{c|}{\textbf{He{\sc i}}}&\multicolumn{1}{c|}{\textbf{Si{\sc ii}}}&\multicolumn{1}{c|}{\textbf{Si{\sc ii}}}\\
\multicolumn{1}{|c|}{\textbf{}}&\multicolumn{1}{c|}{\textbf{}}&\multicolumn{1}{c|}{\textbf{H$\alpha$ EW (\AA)}}&\multicolumn{1}{c|}{\textbf{dd/mm/yr}}&\multicolumn{1}{c|}{\textbf{6678 \AA}}&\multicolumn{1}{c|}{\textbf{6347 \AA}}&\multicolumn{1}{c|}{\textbf{6371 \AA}}\\
\hline
	&		&		&	27/03/17	&	$\times$	&	$\times$	&	$\times$	\\
	&		&		&	29/03/17	&	$\times$	&	$\times$	&	$\times$	\\
	&		&		&	31/03/17	&	$\times$	&	$\times$	&	$\times$	\\
	&		&		&	03/04/17	&	$\times$	&	$\times$	&	$\times$	\\
	&		&		&	06/04/17	&	$\times$	&	$\times$	&	$\times$	\\
	&		&		&	07/04/17	&	$\times$	&	$\times$	&	$\times$	\\
	&		&		&	12/04/17	&	$\times$	&	$\times$	&	$\times$	\\
	&		&		&	13/04/17	&	$\times$	&	$\times$	&	$\times$	\\
	&		&		&	27/04/17	&	$\times$	&	$\times$	&	$\times$	\\
	&		&		&	28/04/17	&	$\times$	&	$\times$	&	$\times$	\\
	&		&		&	30/04/17	&	$\times$	&	$\times$	&	$\times$	\\
	&		&		&	06/05/17	&	$\times$	&	$\times$	&	$\times$	\\
	&		&		&	08/05/17	&	$\times$	&	$\times$	&	$\times$	\\
HD 142926	&	B7IV/V -- B9e	& -3.7 ${\pm0.2}$		&	26/02/16	&	$\times$	&	$\times$	&	$\times$	\\
	&		&		&	12/04/16	&	$\times$	&	$\times$	&	$\times$	\\
	&		&		&	26/04/16	&	$\times$	&	$\times$	&	$\times$	\\
	&		&		&	27/04/16	&	$\times$	&	$\times$	&	$\times$	\\
	&		&		&	22/05/16	&	$\times$	&	$\times$	&	$\times$	\\
	&		&		&	23/05/16	&	$\times$	&	$\times$	&	$\times$	\\
	&		&		&	24/05/16	&	$\times$	&	$\times$	&	$\times$	\\
	&		&		&	25/05/16	&	$\times$	&	$\times$	&	$\times$	\\
	&		&		&	01/02/18	&	$\times$	&	$\times$	&	$\times$	\\
	&		&		&	13/02/19	&	$\times$	&	$\times$	&	$\times$	\\
	&		&		&	23/02/19	&	$\times$	&	$\times$	&	$\times$	\\
	&		&		&	16/03/19	&	$\times$	&	$\times$	&	$\times$	\\
	&		&		&	12/04/19	&	$\times$	&	$\times$	&	$\times$	\\
	&		&		&	15/04/19	&	$\times$	&	$\times$	&	$\times$	\\
	&		&		&	10/05/19	&	$\times$	&	$\times$	&	$\times$	\\
	&		&		&	21/05/19	&	$\times$	&	$\times$	&	$\times$	\\
HD 164447	&	B8Vn	& -7.1 ${\pm0.3}$		&	25/03/16	&	$\times$	&	a	&	a	\\
	&		&		&	20/04/16	&	$\times$	&	a	&	a	\\
	&		&		&	21/04/16	&	$\times$	&	a	&	a	\\
	&		&		&	22/04/16	&	$\times$	&	a	&	a	\\
	&		&		&	22/05/16	&	$\times$	&	a	&	a	\\
	&		&		&	23/05/16	&	$\times$	&	a	&	a	\\
	&		&		&	25/05/16	&	$\times$	&	a	&	a	\\
	&		&		&	30/06/16	&	$\times$	&	a	&	a	\\
	&		&		&	26/03/18	&	$\times$	&	a	&	a	\\
	&		&		&	29/03/18	&	$\times$	&	a	&	a	\\
	&		&		&	14/04/19	&	$\times$	&	a	&	a	\\
	&		&		&	16/04/19	&	$\times$	&	a	&	a	\\
	&		&		&	10/05/19	&	$\times$	&	a	&	a	\\
	&		&		&	19/05/19	&	$\times$	&	a	&	a	\\
HD 171780	&	B5Ve	& -12.2 ${\pm0.3}$		&	20/04/16	&	$\times$	&	$\times$	&	$\times$	\\
	&		&		&	21/04/16	&	$\times$	&	$\times$	&	$\times$	\\
	&		&		&	24/05/16	&	$\times$	&	a	&	a	\\
	&		&		&	25/05/16	&	$\times$	&	a	&	a	\\
	&		&		&	14/04/19	&	$\times$	&	a	&	a	\\
	&		&		&	16/04/19	&	$\times$	&	a	&	a	\\
	&		&		&	12/05/19	&	$\times$	&	a	&	a	\\
	&		&		&	04/11/19	&	$\times$	&	a	&	a	\\
\hline
\end{tabular}}
\end{table*}

\end{document}